\RequirePackage{fix-cm}
\documentclass[twocolumn,epjc3]{svjour3}  
\smartqed  
\usepackage{longtable}
\usepackage{subcaption}
\RequirePackage{graphicx}
\RequirePackage{amsmath}
\usepackage{amssymb}
\usepackage{bigints}
\usepackage{calrsfs}
\DeclareMathAlphabet{\pazocal}{OMS}{zplm}{m}{n}
\newcommand{\La}{\mathcal{L}}
\newcommand{\Lb}{\pazocal{L}}
\usepackage{adjustbox}
\usepackage{multirow}
\usepackage{array}
\usepackage{float}
\usepackage{cite}
\usepackage{balance}

\usepackage[colorlinks=true,linkcolor=blue, citecolor=blue, urlcolor=blue]{hyperref}
\DeclareMathOperator{\sech}{sech}
\journalname{Eur. Phys. J. C}
\begin{document}

\title{Greybody factor and Quasinormal modes of scalar and Dirac field perturbation in Schwarzschild-de Sitter-like black hole in Bumblebee gravity model
}


\author{Yenshembam Priyobarta Singh\thanksref{e1,addr1}
        \and
        Telem Ibungochouba Singh\thanksref{e2,addr1} 
}

\thankstext{e1}{e-mail: priyoyensh@gmail.com     }
\thankstext{e2}{e-mail: ibungochouba@rediffmail.com (corresponding author)}


\institute{Department of Mathematics, Manipur University, Canchipur 795003, India \label{addr1}
}

\date{Received: date / Accepted: date}

\maketitle

\begin{abstract}
In this paper, we study the scalar and Dirac fields perturbation of Schwarzschild-de Sitter-like black hole in bumblebee gravity model. The effective potentials, greybody factors and quasinormal modes of the black hole are investigated by using the Klein-Gordon equation and Dirac equation. To compute the greybody factors for scalar and Dirac fields, we use the general method of rigorous bound.  We also investigate the quasinormal mode using the third order WKB approximation method and P\"{o}schl-Teller fitting method. The impact of Lorentz invariance violation parameter $L$ and cosmological constant $\Lambda$  to the effective potential, greybody factors and quasinormal modes  are  analyzed for different modes. Increasing the parameters $\Lambda$ and $L$ lower the effective potentials and consequently increases the rigorous bound on the greybody factors. Our findings show that the oscillation frequency and the damping rate decrease with increasing $L$. We analyze the Hawking temperature, power spectrum and sparsity of Hawking radiation. Both the peak of power spectrum and total power emitted decrease with increasing $L$. The effect of $L$ on the shadow radius is also discussed.

\end{abstract}

\section{Introduction}
\label{intro}

\par General relativity (GR) describes the gravitation at the classical level and the Standard Model (SM) of particle physics describes particles and the other three fundamental interactions at the quantum level. The unification of these theories will ultimately provide a better understanding of nature.  Several theories of quantum gravity (QG) have been introduced to achieve this unification but the direct test of their properties would be possible only at the Planck scale ($~10^{19}$ GeV) which are currently unavailable, so it is impossible to test these theories directly. However, some signals of QG models are observed at the current low energy scales such as the signals associated with the breaking of Lorentz symmetry \cite{kostelecky1989a,casana2018}.

Several theories give rise to the Lorentz symmetry violation such as noncommutative field theories \cite{mocioiu2000,carroll2001,ferrari2007}, loop quantum gravity theory \cite{gambini1999,ellis2000}, string theory \cite{kostelecky1989b,kostelecky1989c,kostelecky1991}, Einstein-aether theory \cite{jacobson2001,jacobson2004} etc. Moreover, the standard-model extension (SME)  is the effective field theory that relates the SM to GR at a low energy scale and it also shows the violation of Lorentz symmetry at the Planck scale \cite{colladay1997,colladay1998,bluhm2005}. Bumblebee gravity model is the simplest model in which the Lorentz symmetry is spontaneously broken due to the nonzero vacuum expectation value of a single vector field, known as bumblebee field \cite{kostelecky2001,kostelecky2004}.  Bertolami et al. \cite{bertolami2005} derived the  vacuum solutions  for the bumblebee field for purely radial, temporal-radial, and temporal-axial Lorentz symmetry breaking. For the purely radial Lorentz symmetry breaking it is found that the result  corresponds to  new black hole solutions. In 2018, Casana et al. \cite{casana2018} derived an exact Schwarzschild-like black hole solution in the bumblebee gravity model. Following this,   within the framework of the bumblebee gravity model, several static spherically symmetric solutions have been obtained including  the solution with global monopole \cite{gullu2022}, with a Gauss-Bonnet term \cite{ding2022}. Ref. \cite{ovgun2019} derived a traversable wormhole solution in the framework of a bumblebee gravity model. In 2020, Ding et al. \cite{ding2020} claimed that the exact solution of Kerr-like black hole in the bumblebee gravity model was found. However, Refs. \cite{ding2021,maluf2022} revealed that the solution was actually wrong.  Later, it is found that the result can be considered as an approximate solution  of the bumblebee field equation \cite{liu2023}. The actual solution of a slowly rotating black hole in the bumblebee gravity model is derived in \cite{ding2021,kanzi2022}. Moreover, an exact rotating BTZ-like black hole solution  \cite{ding2023} and higher dimensional AdS-like black hole solution \cite{ding2023b} were derived. The solutions of black holes with an effective cosmological constant, which comes from a suitable choice for the bumblebee potential  were obtained in \cite{maluf2021}. The resulting black hole is either Schwarzschild-de Sitter or Schwarzschild-anti-de Sitter-like black hole according to $\Lambda>0$ or $\Lambda<0$.

Recently, the study of bumblebee model has been increased since the black hole structure and surrounding accretion processes may differ from the original models. The violation of Lorentz symmetry leads to the modification of black hole dynamics, including the radiation emission. The potential observing effects of the Lorentz symmetry breaking have been investigated in black hole thermodynamics \cite{kanzi2019,gomes2020,sakalli2023,karmakar2023,priyo2022,onika2022,media2023,onika2023,onika2023b}, gravitational
lensing \cite{ovgun2018,li2020,carvalho2021,mangut2023}, shadow \cite{maluf2021,jha2021,wang2022,vagnozzi2023}, accretion \cite{yang2019}, greybody factors \cite{kanzi2019,kanzi2021,uniyal2023} and quasinormal
modes \cite{liu2023,kanzi2021,oliveira2019,oliveira2021,gogoi2022,chen2023,lin2023}.

Hawking \cite{hawking1975} proposed a thermal radiation  emitted from a black hole known as Hawking radiation. Before reaching an observer and being detected, this radiation has to propagate through curved spacetime geometry. However, the surrounding spacetime acts as a  potential barrier. This barriers act as a filter depending on
the frequencies of the propagating waves. Some parts of the Hawking radiation are reflected back into the black hole, while  some transmitted to spatial infinity. Therefore, there is a deviation from the blackbody radiation spectrum emitted from the black hole’s horizon and the observed radiation. Greybody
factors represent the amount of particles or waves that are
transmitted through the  potential barrier. There are different methods for calculating the greybody factors such as matching technique \cite{fernando2005,kim2008}, WKB approximation method for high gravitational potential \cite{parikh2000,konoplyaa2020} and rigorous bound method \cite{visser1999,boonserm2008,sakalli2022,boonserm2008b,
ngampitipan2013,boonserm2013,boonserm2014a,boonserm2014b,boonserm2018,badawi2024}. In our work, we will use the rigorous bound method to compute the greybody factors. Ref. \cite{lenzi2023} introduce a new semi-analytical method to obtain the greybody factors using Korteweg-de Vries (KdV) integrals. Another interesting aspect to note, is the  sparsity of the Hawking radiation during the evaporation process \cite{gray2016,miao2017,hod2015,hod2016}. The Hawking emission is known to be extremely sparse, that is the average time  between emission of successive Hawking quanta is significantly larger than the characteristic time-scale of individual Hawking emission.


When a black hole is perturbed, the gravitational wave is emitted and the dynamical evolution  can be classified into three distinct stages \cite{konoplya2011}: the first short period of the initial outburst of the wave, the second stage consists of a long period  of  damping (quasinormal) oscillations and the final stage is an asymptotic tails behaviour at very late times. The real parts of the quasinormal frequencies (QNFs) represent the oscillation frequencies of the perturbation  and the imaginary parts are related to the damping \cite{detweiler1980}. There has been a lot of interest in the study of quasinormal modes particularly since the detection of the gravitational waves by LIGO and Virgo \cite{abbott2016}. Quasinormal frequencies can be calculated using different methods such as WKB approximation \cite{schutz1985,iyer1987b,iyer1987a,wahlang2017},   continuous fractions method \cite{leaver1985}, Pösch–Teller fitting method \cite{ferrari1984} and  Frobenius method \cite{konoplya2011}. Exact analytical calculations for quasinormal spectra of black holes is performed in very few cases \cite{cardoso2003,panotopoulos2018,rincon2018}. In this work, we will use the 3rd order WKB method  and Pösch–Teller fitting method  to investigate the effect of bumblee gravity in the quasinormal mode  of scalar and Dirac field perturbations in the background of Schwarzschild-de Sitter-like black hole. Eikonal limit is used to calculate the high frequency quasinormal modes. In this limit, the real part of the quasinormal frequency is related to the unstable circular null geodesics of the black hole \cite{cardoso2009}. The photographs of the supermassive black holes of $M87^*$ and $Sgr A^*$ taken by the Event Horizon Telescope (EHT) \cite{akiyama2019a,akiyama2019b,akiyama2019c,akiyama2019d,akiyama2019e,akiyama2019f,akiyama2022} sparked a lot of interest in the study of black hole shadows. These findings ignited interest in both GR and modified gravity theory,  and it also provides a new opportunity to observe black holes and test gravity theories in the strong regime. In our work, we will investigate the effect of bumblebee gravity on the shadow of Schwarzschild-de Sitter-like black hole.  

The paper is organized as follows: A brief discussion of Schwarzschild-de Sitter-like black hole is given in Section \ref{SdS}. In Section \ref{section scalar}, we study the scalar field perturbation using Klein-Gordon equation. Section \ref{section dirac} is devoted to the analysis of Dirac field perturbation using Dirac equation within the framework of Newman-Penrose formalism. Next, we compute the greybody factors for both scalar and Dirac fields in Section \ref{section greybody}. The quasinormal frequencies of scalar and Dirac field perturbation are discussed in Section \ref{section quasi}. In Section \ref{sparsity}, we examine the effect of bumblebee gravity on power spectrum and sparsity of Hawking radiation. Section \ref{geodesic}, deals with the null geodesic and the shadow of the black hole. In Section \ref{eikonal}, we discuss how quasinormal is linked with shadow of the black hole. In Section \ref{sec constraint}, we try to constrain parameters using observables data from EHT. Lastly, conclusion is given in Section \ref{conclusion}. 
 
\section{Schwarzschild-de Sitter-like black hole in Bumblebee gravity model }\label{SdS}
\begin{figure}[h!]
\centering
  \centerline{\includegraphics[height=160pt,width=230pt]{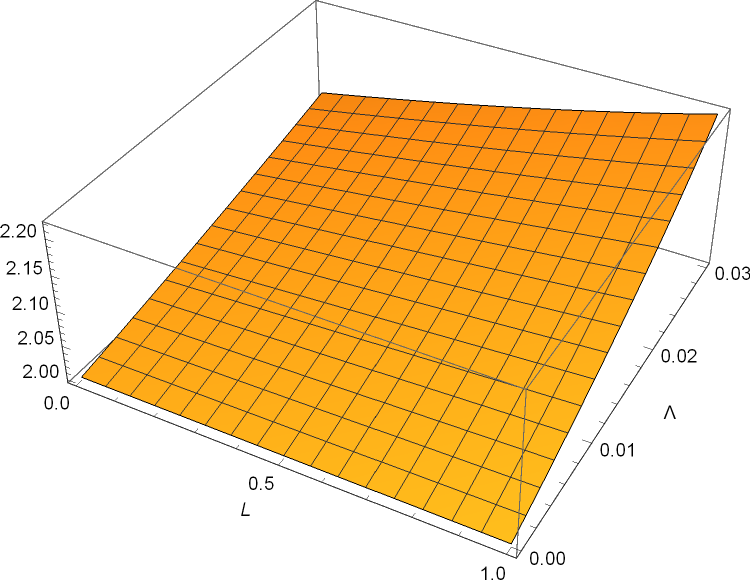}}
  \caption{Relations between the event horizon $r_h$ and the metric parameters. Here we set $M=1$.  } 
  \label{fig event}
\end{figure}

\begin{figure}[h!]
\centering
  \centerline{\includegraphics[height=160pt,width=230pt]{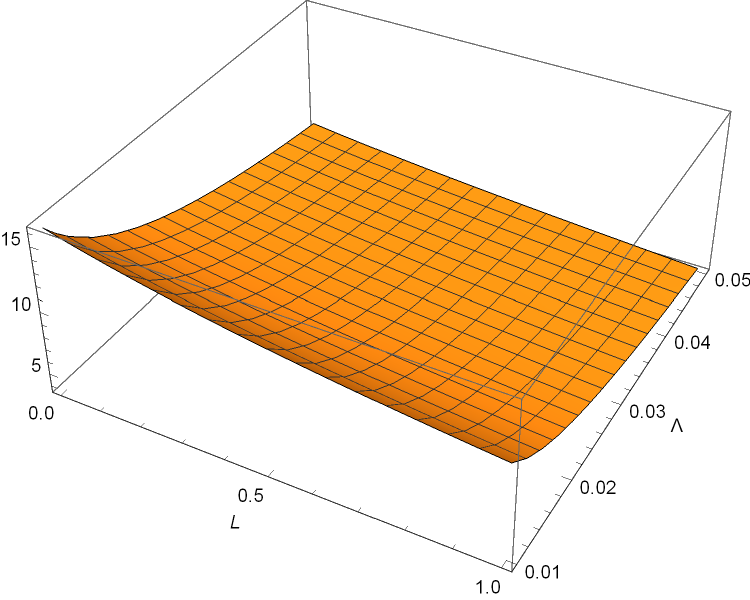}}
  \caption{Relations between the cosmological horizon $r_c$ and the metric parameters. Here we set $M=1$. } 
  \label{fig cosmological}
\end{figure}

The line element of Schwarzschild-de Sitter-like black hole in Bumblebee gravity model is given by \cite{maluf2021}

\begin{align} \label{metric}
ds^2=f(r) dt^2-\dfrac{(1+L)}{f(r)} dr^2
-r^2 d\theta^2 -r^2 \sin^2\theta d\phi^2,
 \end{align}
where the function $f(r)$ has the following form 
\begin{align}
f(r)=1-\dfrac{2M}{r}-(1+L)\dfrac{\Lambda}{3} r^2,
\end{align}
in which, $M$, $\Lambda$ and $L$  are the mass of the black hole, cosmological constant and Lorentz violating parameter respectively.  It is readily seen that, in the limit $L\rightarrow 0$, one can see that the original Schwarzschild-de Sitter black hole is recovered and when $\Lambda=0$, one can get the Schwarzschild-like black hole in bumblebee gravity model. The horizons of the Schwarzschild-de Sitter-like black hole are the real positive roots of the  equation 
\begin{align}
f(r)=\dfrac{(1+L)\Lambda}{3 r} \left(r-r_h\right) \left(r_c-r\right) \left(r-r_-\right)=0.
\end{align} 
If $-1<L<\dfrac{1-9 \Lambda M^2}{9 \Lambda M^2}$, we get two horizons namely  event horizon $r_h$ and the cosmological horizon $r_c$  which are given by
\begin{align}\label{horizon}
r_h=\dfrac{2 M}{\sqrt{3 \Pi}} \cos \left(\dfrac{\pi+\Phi}{3}\right), \\
r_c=\dfrac{2 M}{\sqrt{3 \Pi}} \cos \left(\dfrac{\pi-\Phi}{3}\right),
\end{align}
where $\Phi=\cos^{-1} \left( 3 \sqrt{\Pi} \right)$ and $\Pi=(1+L)\Lambda M^2$ \cite{rahman2012}.
 Further, the two horizons coincide when $L=(1-9 \Lambda M^2)/(9 \Lambda M^2)$ and the region $r_h<r<r_c$ is the domain of outer communication. From, Figs. \ref{fig event} and \ref{fig cosmological}, one can see that the event horizon radius increases with increasing the Lorentz-violation parameter $L$ and  cosmological constant $\Lambda$. However, the increase in $L$ and $\Lambda$ reduce the cosmological horizon.

\section{Scalar Perturbation}\label{section scalar}

In this section, we will discuss the scalar perturbation  in the geometry of the Schwarzschild-de Sitter-like black hole. The effective potential of the scalar perturbation will be derived by using the Klein-Gordon equation \cite{fulling1989}
\begin{align}\label{klein}
\dfrac{1}{\sqrt{-g}} \partial_\mu \left(\sqrt{-g} ~g^{\mu \nu} \partial_\nu \Psi \right)-m^2 \Psi=0,
\end{align}
where $g_{\mu\nu}$, $g^{\mu \nu}$, $g$ and $m$ are the metric tensor, inverse metric tensor, determinant of the metric tensor and mass of the scalar particle respectively. The determinant of the spacetime metric \eqref{metric} is found to be 
\begin{align}\label{det}
g=-\left(1+L\right) r^4 \sin^2\theta.
\end{align}
Using Eqs. \eqref{metric} and \eqref{det} in Eq. \eqref{klein}, we obtain
\begin{align}\label{klein3}
&\dfrac{1}{f} \partial^{2}_{t} \Psi-\dfrac{1}{r^2 \left(1+L\right)} \left(2 r f \partial_r \Psi +r^2 f' \partial_r \Psi+r^2 f ~\partial^{2}_{r} \Psi \right) \nonumber\\
&-\dfrac{1}{r^2 \sin\theta} \left(\cos\theta~ \partial_{\theta}\Psi +\sin\theta \partial^{2}_\theta \Psi \right)-\dfrac{1}{r^2 \sin^2\theta } \partial^{2}_{\phi} \Psi \nonumber\\
&-m^2 \Psi=0.
\end{align} 
Eq. \eqref{klein3} contains the variables $t$, $r$, $\theta$ and $\phi$. To study the effective potential for the tunneling of scalar particle, we define the scalar field as
\begin{align} \label{klein4}
\Psi=R(r) A(\theta) \exp\left(-i \omega t\right) \exp \left(i \widetilde{m} \phi\right),
\end{align}
where $\omega$ and $\widetilde{m}$ represent the energy of the particle and the azimuthal quantum number respectively.  Now, Eq. \eqref{klein3} reduces to
\begin{align}\label{klein5}
& r^2\left[-\dfrac{\omega^2}{f}-\dfrac{1}{R \left(1+L\right)} \left(\dfrac{2 f}{r} R'+f'R'+f R''\right)-m^2 \right]	\nonumber\\
& -\dfrac{1}{A\sin\theta} \left(\cos\theta A'+\sin\theta A''-\dfrac{\widetilde{m}^2 A}{\sin\theta} \right)=0.
\end{align} 
Changing the independent variable $\theta$ to $\cos^{-1}z$, the angular part of Eq. \eqref{klein5} is derived as
\begin{align}\label{klein6}
\left(1-z^2\right) A''-2z A'- \left(\lambda_s +\dfrac{\widetilde{m}^2}{1-z^2}\right)A=0,
\end{align}
where $\lambda_s$ denotes the eigenvalue. If $\lambda_s=-l(1+l)$, one can see that the angular equation is simply the Legendre differential equation. The radial part of Eq. \eqref{klein5} becomes
\begin{align}\label{klein7}
&R''+\left(\dfrac{2}{r}+\dfrac{f'}{f} \right)R'+\left\lbrace \dfrac{\omega^2 \left(1+L\right)}{f^2}+\dfrac{\lambda_s \left(1+L\right)}{f r^2}   \right. \nonumber \\ &\left.  +\dfrac{m^2 \left(1+L\right)}{f} \right\rbrace R=0.
\end{align}

To obtain a one-dimensional Schr\"{o}dinger-like wave equation, we use the transformation $R=u/r$, together with the tortoise coordinate $r_*$ defined by
\begin{align}
\dfrac{dr_*}{dr}=\dfrac{\sqrt{1+L}}{f}.
\end{align}

Thus, Eq. \eqref{klein7} can be  reduced to a one-dimensional Schr\"{o}dinger-like wave equation as
\begin{align}\label{klein9}
\dfrac{du^2}{dr_{*}^2}+\left(\omega^2-V_{s}\right)u=0.
\end{align}
Here $V_{s}$ is the effective potential of massive scalar particle which is given by
\begin{align}\label{klein10}
V_{s}=&f\left( \dfrac{f'}{r\left(1+L\right)}+\dfrac{l(l+1)}{r^2}+m^2\right)\nonumber\\
=&  \left(1-\dfrac{2M}{r}-(1+L)\dfrac{\Lambda}{3} r^2\right) \nonumber\\ & \times \left( \dfrac{2M}{(1+L)r^3}-\dfrac{2\Lambda}{3} +\dfrac{l(l+1)}{r^2}+m^2\right).
\end{align}

The effective potential depends on the mass of the black hole $M$, cosmological constant $\Lambda$, Lorentz violation parameter $L$, mass of the scalar particle $m$ and $l$. If $m=0$, Eq. \eqref{klein} reduces to the effective potential of massless scalar particle for Schwarzschild-de Sitter-like black hole. For $m=0$ and $\Lambda=0$, the effective potential of massless scalar particles for for Schwarzschild-like black hole \cite{kanzi2019}.  If $L=0$, we recover the effective potential of massive scalar particles for Schwarzschild-de Sitter black hole \cite{toshmatov2017,zinhailo2024}. Further, if $L=0$ and $m=0$, we get the effective potential of massless scalar particles for Schwarzschild-de Sitter black hole \cite{zhidenko2004}.   It is worth mentioning that if $m=0$, $\Lambda=0$ and $L=0$, then $V_s$ reduces to the effective potential of Schwarzschild black hole \cite{iyer1987b}. 
\begin{center}
\begin{figure*}[h!]
 \centering
  \begin{subfigure}[b]{0.32\textwidth}
    \centering
    \includegraphics[height=150pt,width=150pt]{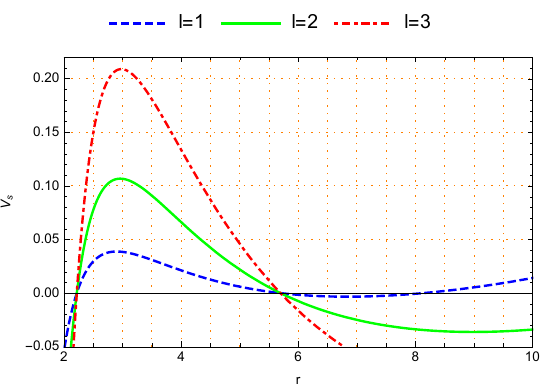} 
    \caption{}
    \label{fig vsl}
  \end{subfigure}
  \hfill
  \begin{subfigure}[b]{0.32\textwidth}
    \centering
    \includegraphics[height=150pt,width=150pt]{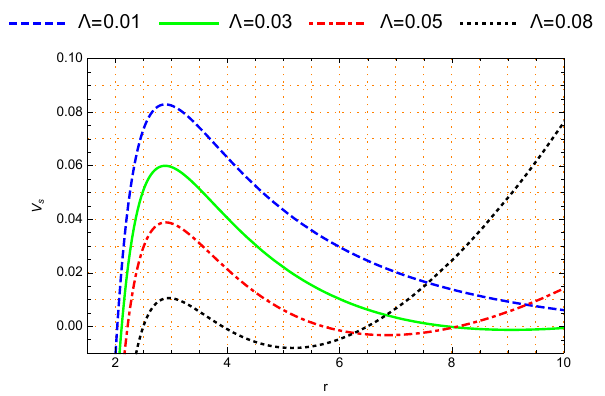} 
    \caption{}
    \label{fig vsL}
  \end{subfigure}
   \hfill
  \begin{subfigure}[b]{0.32\textwidth}
    \centering
    \includegraphics[height=150pt,width=150pt]{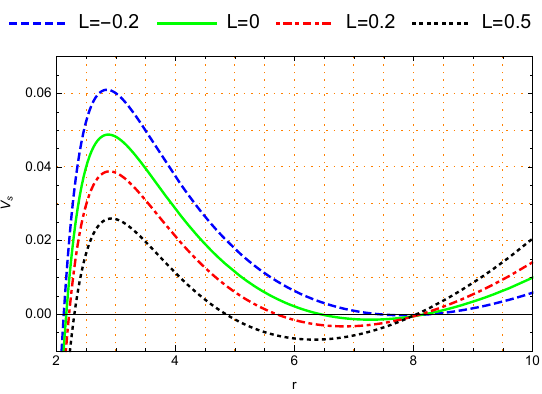} 
    \caption{}
    \label{fig vsLL}
    \end{subfigure}
  \caption{Variation of the effective potential for the massless scalar field: (a) for different values of $l$ with fixed $M=1$, $\Lambda=0.05$ and $L=0.2$; (b) for different values of $\Lambda$ with fixed $M=1$, $l=1$ and $L=0.2$; (c) for different values of $L$ with fixed $M=1$, $\Lambda=0.05$ and $l=1$. }
  \label{fig vs}
\end{figure*}
\end{center}
\begin{center}
\begin{figure*}[h!]
 \centering
  \begin{subfigure}[b]{0.32\textwidth}
    \centering
    \includegraphics[height=150pt,width=150pt]{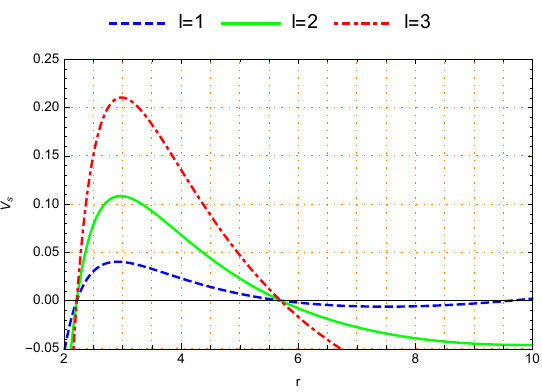} 
    \caption{}
    \label{fig vsml}
  \end{subfigure}
  \hfill
  \begin{subfigure}[b]{0.32\textwidth}
    \centering
    \includegraphics[height=150pt,width=150pt]{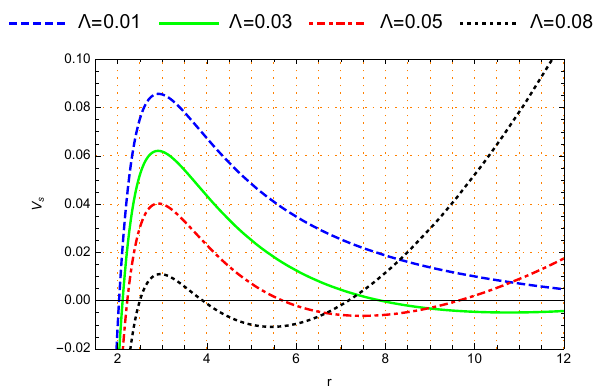} 
    \caption{}
    \label{fig vsmL}
  \end{subfigure}
   \hfill
  \begin{subfigure}[b]{0.32\textwidth}
    \centering
    \includegraphics[height=150pt,width=150pt]{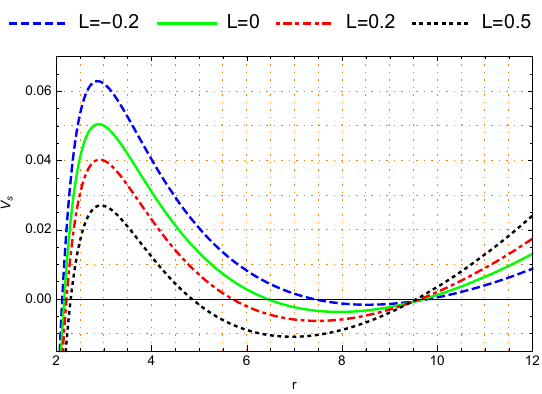} 
    \caption{}
    \label{fig vsmLL}
    \end{subfigure}
 \caption{Variation of the effective potential for the massive scalar field: (a) for different values of $l$ with fixed $M=1$, $m=0.1$, $\Lambda=0.05$ and $L=0.2$; (b) for different values of $\Lambda$ with fixed $M=1$, $m=0.5$,  $l=1$ and $L=0.2$; (c) for different values of $L$ with fixed $M=1$, $m=0.1$,  $\Lambda=0.05$ and $l=1$. }
  \label{fig vsm}
\end{figure*}
\end{center}
The behaviour of the effective potentials of massless and massive scalar particles for different values of $l$, $\Lambda$ and $L$ are illustrated in  Figs. \ref{fig vs} and \ref{fig vsm} respectively.  The blue dashed line in Figs. \ref{fig vsLL} and \ref{fig vsmLL} represent the effective potentials of massless and massive scalar particles in the classical Schwarzschild-de Sitter black hole.  From Figs. \ref{fig vsl} and \ref{fig vsml}, it is observed that the effective potentials increase with increasing $l$. It is also noticed  that  the peak value of the effective potentials decrease with increasing the parameters $\Lambda$ and $L$ which indicates that the greybody factors will increase with increasing the parameters $\Lambda$ and $L$. Thus, the Lorentz invariance violation plays an important role in reducing the local maximum of the effective potential. Further, for the massless case the position of the peak of the effective potential shifts along the right as $l$, $\Lambda$ and $L$ increase but for massive case, it shifts towards the left.  It is worth noting that the effective potential vanishes at three points namely event horizon ($r_h$), cosmological horizon ($r_c$) and an additional zero point at $r_e$, located above the cosmological horizon. The effective potential has a barrier-like form between $r_h$ and $r_c$, and it is negative between the $r_c$ and $r_e$.

\section{Dirac Perturbation}\label{section dirac}
To find the massless and massive Dirac fields propagating in the geometry of the Schwarzschild-de Sitter-like black hole, we shall use the Newman-Penrose (NP) formalism \cite{newman1962}.  The null tetrad vectors for the Schwarzschild-de Sitter-like black hole are defined as
\begin{align}\label{null1}
&l^\mu=\left\lbrace f^{-1}, \dfrac{1}{\sqrt{1+L}},0,0		\right\rbrace, \nonumber\\
&n^\mu=\left\lbrace \dfrac{1}{2}, \dfrac{-f}{2\sqrt{1+L}},0,0		\right\rbrace, \nonumber\\
&m^\mu=\left\lbrace	0,0, \dfrac{1}{\sqrt{2}r}, \dfrac{i \csc\theta}{\sqrt{2}}		\right\rbrace,\nonumber\\
&\overline{m}^\mu=\left\lbrace	0,0, \dfrac{1}{\sqrt{2}r}, \dfrac{-i \csc\theta}{\sqrt{2}}		\right\rbrace.
\end{align}
The dual co-tetrad vectors of Eq. \eqref{null1} are given by
\begin{align}\label{null2}
&l_\mu=\left\lbrace	1, \dfrac{-\sqrt{1+L}}{f}, 0,0 \right\rbrace, \nonumber\\
&n_\mu=\left\lbrace	\dfrac{f}{2}, \dfrac{\sqrt{1+L}}{2}, 0,0 \right\rbrace, \nonumber\\
&m_\mu=\left\lbrace	0,0,  \dfrac{-r}{\sqrt{2}}, \dfrac{-i r\sin\theta}{\sqrt{2}} \right\rbrace, \nonumber\\
&\overline{m}_\mu=\left\lbrace	0,0,  \dfrac{-r}{\sqrt{2}}, \dfrac{i r\sin\theta}{\sqrt{2}} \right\rbrace.
\end{align}
The non vanishing spin coefficients of the black hole are found to be

\begin{align}\label{spin}
&\rho=\dfrac{-1}{r\sqrt{1+L} }, \quad \mu=\dfrac{-f}{2 r \sqrt{1+L}}, \quad \gamma=\dfrac{f'}{4\sqrt{1+L}}, \nonumber\\
&\beta=\dfrac{\cot\theta}{2\sqrt{2} r}, \quad \alpha=-\dfrac{\cot\theta}{2\sqrt{2} r}.
\end{align}
The Chandrasekhar-Dirac equations \cite{chandrasekhar} in the NP formalism are given by  
\begin{align}\label{CDE}
&\left(	D+\epsilon-\rho \right) F_1+\left(	\bar{\delta}+\pi -\alpha\right) F_2=i \mu^* G_1, \nonumber\\
&\left( \delta+\beta-\tau \right) F_1+ \left(\Delta+\mu -\gamma \right) F_2=i \mu^* G_2, \nonumber\\
&\left(D+\bar{\epsilon}-\bar{\rho} 	\right) G_2 -\left( \delta+\bar{\pi}-\bar{\alpha} \right) G_1=i \mu^* F_2, \nonumber\\
& \left(\Delta+\bar{\mu}-\bar{\gamma} \right) G_1- \left(\bar{\delta}+\bar{\beta}-\bar{\tau} \right)G_2=i \mu^* F_1,
\end{align}
where $F_1$, $F_2$, $G_1$ and $G_2$ represent the Dirac spinors. $D$, $\Delta$, $\delta$ and $\bar{\delta}$ are the directional derivatives defined by
\begin{align}\label{operator}
& D=l^\mu \dfrac{\partial}{\partial x^\mu} = f^{-1} \dfrac{\partial}{\partial t}+\dfrac{1}{\sqrt{1+L}} \dfrac{\partial}{\partial r}, \nonumber\\
&\Delta=n^\mu \dfrac{\partial}{\partial x^\mu} =\dfrac{1}{2} \dfrac{\partial}{\partial t}-\dfrac{f}{2\sqrt{1+L}} \dfrac{\partial}{\partial r}, \nonumber\\
& \delta= m^\mu \dfrac{\partial}{\partial x^\mu} = \dfrac{1}{\sqrt{2}r} \dfrac{\partial}{\partial \theta}+\dfrac{i \csc \theta}{\sqrt{2}r} \dfrac{\partial}{\partial \phi},		\nonumber\\
& \bar{\delta}= \overline{m}^\mu \dfrac{\partial}{\partial x^\mu} =\dfrac{1}{\sqrt{2} r} \dfrac{\partial}{\partial \theta}-\dfrac{i \csc \theta}{\sqrt{2}r} \dfrac{\partial}{\partial \phi}.
\end{align}

To solve the Chandrasekhar-Dirac equations  \eqref{CDE}, we will consider the spinor  in the form of \\$F=F(r,\theta) \exp[i(\omega{}t+\tilde{m} {}\phi)]$, where $\omega$ is the frequency of the wave corresponding to the Dirac particle and $\tilde{m}$ is the azimuthal quantum number of the wave. To get radial and angular parts of Eq. \eqref{CDE}, we choose
\begin{align}\label{spinor}
&F_1=f_{1}(r) A_{1}(\theta) \exp[i(\omega {}t+\tilde{m}{}\phi)], \nonumber\\
&G_1=g_{1}(r) A_{2}(\theta) \exp[i(\omega {}t+\tilde{m}{}\phi)], \nonumber\\
&F_2=f_{2}(r) A_{3}(\theta) \exp[i(\omega {}t+\tilde{m}{}\phi)], \nonumber\\
&G_2=g_{2}(r) A_{4}(\theta) \exp[i(\omega {}t+\tilde{m}{}\phi)].
\end{align}

Substituting the spin coefficients \eqref{spin}, directional derivatives \eqref{operator} and the spinors \eqref{spinor} in Eq. \eqref{CDE}, we obtain

\begin{align}\label{CDE2}
&A_1 \left[ \dfrac{1}{\sqrt{1+L}}+\dfrac{i r \omega}{f}+\dfrac{r}{\sqrt{1+L}} \dfrac{d}{d r}\right] f_1+\dfrac{1}{\sqrt{2}} f_2 ~\tilde{L}^+ A_3	\nonumber\\
&=	i \mu^* r A_2 g_1,  \nonumber\\
& f A_3 \left[\dfrac{1}{\sqrt{1+L}}+\dfrac{f' r}{2f \sqrt{1+L}}-\dfrac{i r \omega}{f}+\dfrac{r}{\sqrt{1+L}} \dfrac{d}{d r}\right] f_2 \nonumber\\
& -\sqrt{2} f_1 \tilde{L} A_1=-2 i \mu^* r A_4 g_2, \nonumber\\
& A_4 \left[ \dfrac{1}{\sqrt{1+L}}+\dfrac{i r \omega}{f}+\dfrac{r}{\sqrt{1+L}} \dfrac{d}{d r}\right] g_2-\dfrac{1}{\sqrt{2}} g_1 ~\tilde{L} A_2	\nonumber\\
&=	i \mu^* r A_3 f_2,  \nonumber\\
& f A_2 \left[\dfrac{1}{\sqrt{1+L}}+\dfrac{f' r}{2f \sqrt{1+L}}-\dfrac{i r \omega}{f}+\dfrac{r}{\sqrt{1+L}} \dfrac{d}{d r}\right] g_1 \nonumber\\
& +\sqrt{2} g_2 \tilde{L}^+ A_4=-2 i \mu^* r A_1 f_1,
\end{align}
where $\tilde{L}$ and $\tilde{L}^+$ are the angular operators, defined by
\begin{align}\label{angularoperator}
&\tilde{L}^+= \dfrac{d}{d\theta}+\dfrac{\tilde{m}}{\sin\theta}+\dfrac{\cot\theta}{2},\nonumber\\
&\tilde{L}= \dfrac{d}{d\theta}-\dfrac{\tilde{m}}{\sin\theta}+\dfrac{\cot\theta}{2}.
\end{align}

By considering, $f_1=g_2$, $f_2=g_1$, $A_1=A_2$ and $A_3=A_4$, from Eqs. \eqref{CDE2} and \eqref{angularoperator}, the radial and the angular parts are

\begin{align}\label{CDE3}
& \left[ \dfrac{1}{\sqrt{1+L}}+\dfrac{i r \omega}{f}+\dfrac{r}{\sqrt{1+L}} \dfrac{d}{d r}\right] g_2-	i \mu^* r  g_1 =-\lambda_1 g_1, 	 \nonumber\\
& f \left[\dfrac{1}{\sqrt{1+L}}+\dfrac{f' r}{2f \sqrt{1+L}}-\dfrac{i r \omega}{f}+\dfrac{r}{\sqrt{1+L}} \dfrac{d}{d r}\right] g_1 \nonumber\\
& +2 i \mu^* r  g_2 =\lambda_2 g_2 ,\nonumber\\
& \left[ \dfrac{1}{\sqrt{1+L}}+\dfrac{i r \omega}{f}+\dfrac{r}{\sqrt{1+L}} \dfrac{d}{d r}\right] g_2-	i \mu^* r  g_1 =\lambda_3 g_1, 	 \nonumber\\
& f \left[\dfrac{1}{\sqrt{1+L}}+\dfrac{f' r}{2f \sqrt{1+L}}-\dfrac{i r \omega}{f}+\dfrac{r}{\sqrt{1+L}} \dfrac{d}{d r}\right] g_1 \nonumber\\
& +2 i \mu^* r  g_2 =-\lambda_4 g_2 
\end{align}
and
\begin{align}\label{angularoperator2}
&\tilde{L}^+ A_3=\sqrt{2} A_1 \lambda_1, ~~~\tilde{L} A_1=\dfrac{1}{\sqrt{2}} A_3 \lambda_2,\nonumber\\
& \tilde{L} A_1=\sqrt{2} A_3 \lambda_3, ~~~\tilde{L}^+ A_3=\dfrac{1}{\sqrt{2}} A_1 \lambda_4,
\end{align}
respectively where  $\lambda_1$, $\lambda_2$, $\lambda_3$ and $\lambda_4$ are the separation constants. To obtain the radial and the angular pair equation, the separation constants are taken as
\begin{align*}
\sqrt{2} \lambda_1=\dfrac{1}{\sqrt{2}}\lambda_4=-\lambda,\dfrac{1}{\sqrt{2}} \lambda_2=\sqrt{2} \lambda_3=\lambda.
\end{align*}

Eqs. \eqref{CDE3} and \eqref{angularoperator2} reduce to 
\begin{align}
&\dfrac{r}{\sqrt{1+L}} \left[\dfrac{d}{dr}+\dfrac{i\omega \sqrt{1+L}}{f}+\dfrac{1}{r} \right] g_2= \left(\lambda+i \mu_* r\right) g_1,  \label{radial1}\\
&\dfrac{f r}{\sqrt{1+L}} \left[ \dfrac{d}{dr}-\dfrac{i \omega \sqrt{1+L}}{f}+\dfrac{1}{r}+\dfrac{f'}{2f} \right] g_1 \label{radial2} \nonumber\\
&=\left(\lambda- i \mu_* r\right)g_2, \\
&
\tilde{L}^+ A_3=-\lambda~ A_1 ,\label{angular1} \\
&\tilde{L} A_1=\lambda ~A_3, \label{angular2}
\end{align}

where $\mu_*$ is the normalized rest mass of the spin-1/2 particle. Angular equations  \eqref{angular1} and \eqref{angular2} lead to the spin-weighted spheroidal harmonics \cite{newman1962,goldberg1967} with the following eigenvalue
\begin{align}
\lambda=-\left(l+\dfrac{1}{2}\right).
\end{align}
Taking $g_{1}(r)=\dfrac{\widetilde{\Psi}_1}{r}$ and $g_{2}(r)=\dfrac{\widetilde{\Psi}_2}{r}$, the radial equations \eqref{radial1} and \eqref{radial2},  reduce to
\begin{align}
&\dfrac{1}{\sqrt{1+L}} \left[\dfrac{d}{d r}+ \dfrac{i \omega \sqrt{1+L}}{f} \right] \widetilde{\Psi}_2= \left(\dfrac{\lambda}{r}+i \mu_* \right) \widetilde{\Psi}_1, \label{radial3} \\
&\dfrac{f}{\sqrt{1+L}} \left[ \dfrac{d}{dr}-\dfrac{i \omega \sqrt{1+L}}{f} +\dfrac{f'}{2 f} \right] \widetilde{\Psi}_1=  \left(\dfrac{\lambda}{r}-i \mu_* \right) \widetilde{\Psi}_2.\label{radial4}
\end{align}

Further making the following transformations
$
\widetilde{\Psi}_1= f^{-\frac{1}{2}} R_{1} (r), ~~~ \widetilde{\Psi}_2=R_{2} (r)
$
and using the tortoise coordinate ($r_*$) defined by
\begin{align}
 \dfrac{d}{dr_*}=\dfrac{f}{\sqrt{1+L}} \dfrac{d}{dr},
\end{align}
Eqs. \eqref{radial3} and \eqref{radial4} transform to
\begin{align}
&\left(\dfrac{d}{dr_*}+i \omega \right) R_2= \sqrt{f} \left( \dfrac{\lambda}{r}+i \mu_* \right) R_1, \label{radial5} \\
& \left( \dfrac{d}{d r_*} -i \omega \right) R_1= \sqrt{f} \left( \dfrac{\lambda}{r}-i \mu_* \right) R_2.  \label{radial6}
\end{align}

\subsection{Massless Dirac particle}
For massless Dirac particle (neutrino), Eqs. \eqref{radial5} and \eqref{radial6} reduce to
\begin{align}\
&\left(\dfrac{d}{dr_*}+i \omega \right) R_2=  \lambda \dfrac{\sqrt{f}~}{r} R_1, \label{massless1} \\
& \left( \dfrac{d}{d r_*} -i \omega \right) R_1= \lambda  \dfrac{\sqrt{f}~}{r} R_2.  \label{massless2}
\end{align}
Eqs. \eqref{massless1} and \eqref{massless2} can be combined by letting
\begin{align}
Z_+=R_1 +R_2,\\
Z_-=R_1-R_2.
\end{align}
Thus, Eqs. \eqref{massless1} and \eqref{massless2} can be written as
\begin{align}
\left( \dfrac{d}{d r_*}-\lambda \dfrac{\sqrt{f}}{r} \right) Z_+= i \omega Z_-, \label{massless3}\\
\left( \dfrac{d}{d r_*}+\lambda \dfrac{\sqrt{f}}{r} \right) Z_-= i \omega Z_+.\label{massless4}
\end{align}

\begin{center}
\begin{figure*}[]
 \centering
  \begin{subfigure}[b]{0.32\textwidth}
    \centering
    \includegraphics[height=150pt,width=150pt]{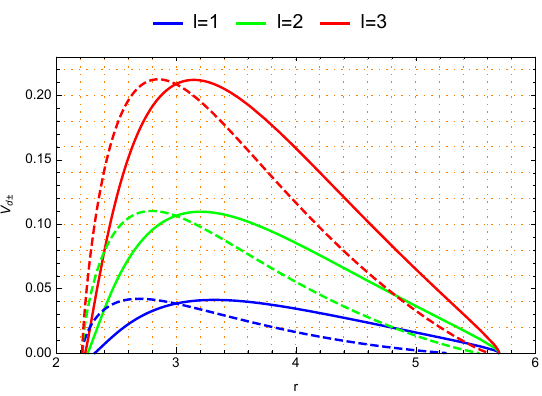} 
    \caption{}
    \label{fig vdl}
  \end{subfigure}
  \hfill
  \begin{subfigure}[b]{0.32\textwidth}
    \centering
    \includegraphics[height=150pt,width=150pt]{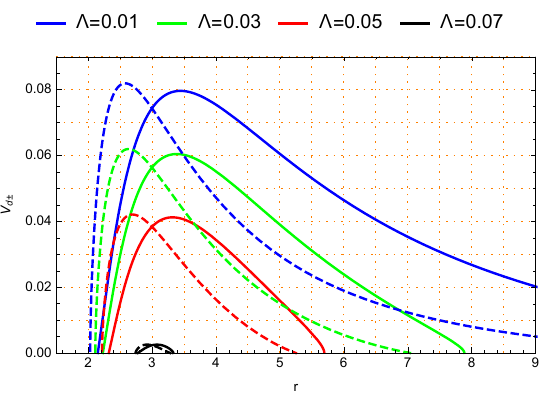} 
    \caption{}
    \label{fig vdL}
  \end{subfigure}
   \hfill
  \begin{subfigure}[b]{0.32\textwidth}
    \centering
    \includegraphics[height=150pt,width=150pt]{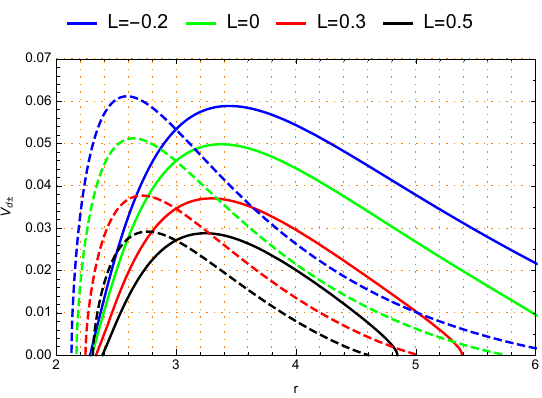} 
    \caption{}
    \label{fig vdLL}
    \end{subfigure}
  \caption{Variation of the effective potential for the massless Dirac field: (a) for different values of $l$ with fixed $M=1$, $\Lambda=0.05$ and $L=0.2$; (b) for different values of $\Lambda$ with fixed $M=1$, $l=1$ and $L=0.2$; (c) for different values of $L$ with fixed $M=1$, $\Lambda=0.05$ and $l=1$. The solid  and dashed line represent ${V_d}_+$ and ${V_d}_-$ respectively. }
  \label{fig vd}
\end{figure*}
\end{center}

From Eqs. \eqref{massless3} and \eqref{massless4}, we obtain the pair of one dimensional Schr\"{o}dinger-like wave equations 
\begin{align}
\left(\dfrac{d^2}{d r_{*}^{2} }+ \omega^2 \right) Z_+={V_{d}}_+ Z_+ \label{waved}\\
\left(\dfrac{d^2}{d r_{*}^{2} }+ \omega^2 \right) Z_-={V_{d}}_-Z_-,
\end{align}
where ${V_{d}}_\pm$ are the effective potentials for the massless Dirac field:
\begin{align}\label{vmasslessdirac}
{V_{d}}_\pm&=\dfrac{\lambda^2 f}{r^2} \pm \lambda \dfrac{d}{d r_*} \left(\dfrac{\sqrt{f}}{r} \right)		\nonumber\\
&=\dfrac{\lambda^2 f}{r^2} \pm \dfrac{\lambda \sqrt{f} \left(3M-r\right)}{r^3 \sqrt{1+L}}.
\end{align}
It is evident from Eq. \eqref{vmasslessdirac} that similar to the scalar field's effective potential, the effective potential of massless Dirac fields  depend on $M$, $\Lambda$, $l$, and $L$. In the absence of Lorentz violation parameter $L$,  ${V_{d}}_\pm$ reduce to the effective potentials of Schwarzschild-de Sitter black hole for Dirac field and the result is consistent with the literature \cite{zhidenko2004}.
 To investigate the variation of effective potential caused by $l$, $\Lambda$ and $L$, we plot the effective potential ${V_{d}}_+$ of massless Dirac field for different values of $l$, $\Lambda$ and $L$ in Fig. \ref{fig vd}. One can see that the peak of the effective potential increases as $l$ increases and the position of the  peak of ${V_d}_+$ moves along the left  but for ${V_d}_-$ it moves towards right. Further with increasing $\Lambda$ and $L$, the effective potential decreases and the position of the peaks of ${V_d}_+$ and ${V_d}_-$ shift towards left and right respectively.  Further, the position of the peak of ${V_d}_-$ occurs at a lower value of $r$ as compare to that of ${V_d}_+$.

\begin{center}
\begin{figure*}[h!]
 \centering
  \begin{subfigure}[b]{0.32\textwidth}
    \centering
    \includegraphics[height=150pt,width=150pt]{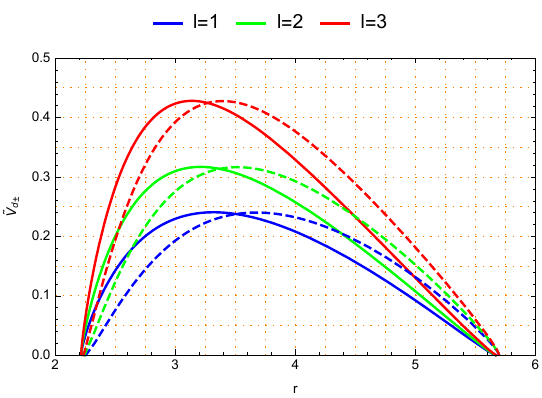} 
    \caption{}
    \label{fig vdml}
  \end{subfigure}
  \hfill
  \begin{subfigure}[b]{0.32\textwidth}
    \centering
    \includegraphics[height=150pt,width=150pt]{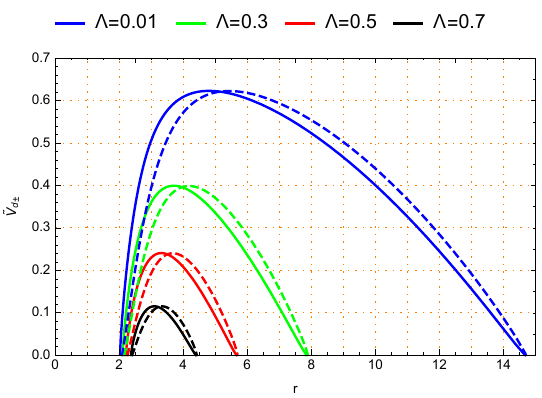} 
    \caption{}
    \label{fig vdmLambda}
  \end{subfigure}
   \hfill
  \begin{subfigure}[b]{0.32\textwidth}
    \centering
    \includegraphics[height=150pt,width=150pt]{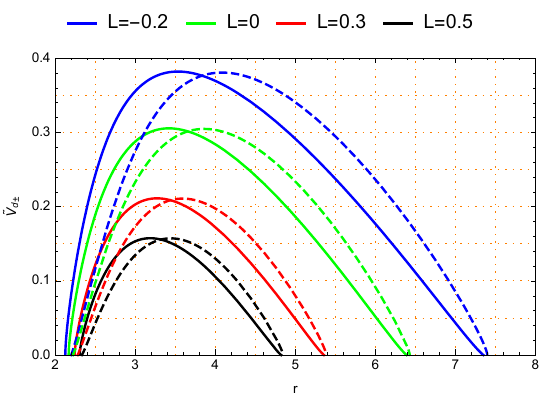} 
    \caption{}
    \label{fig vdmLL}
    \end{subfigure}
  \caption{Variation of the effective potential for the massive Dirac field: (a) for different values of $l$ with fixed $M=1$, $\Lambda=0.05$ and $L=0.2$; (b) for different values of $\Lambda$ with fixed $M=1$, $l=1$ and $L=0.2$; (c) for different values of $L$ with fixed $M=1$, $\Lambda=0.05$ and $l=1$. The solid  and dashed line represent $\widetilde{V_d}_+$ and $\widetilde{V_d}_-$ respectively. }
  \label{fig vdm}
\end{figure*}
\end{center}


\subsection{Massive Dirac field}
To  combine Eqs. \eqref{radial5} and \eqref{radial6}, we take the transformation
\begin{align}
Z=\tan^{-1} \left(\dfrac{\mu_* r}{\lambda}\right).
\end{align}
After some calculations,  Eqs. \eqref{radial5} and \eqref{radial6} reduce to
\begin{align}
\left(\dfrac{d}{d r_*}+i \omega \right) R_2=&\dfrac{\sqrt{f}}{r} \left(\lambda^2+\mu_{*}^2 r^2\right)^{\frac{1}{2}} R_1  \nonumber\\
&\exp\left[ i \tan^{-1} \left(\dfrac{\mu_* r}{\lambda}\right)\right], \label{radial7}\\
\left(\dfrac{d}{d r_*}-i \omega \right) R_1=&\dfrac{\sqrt{f}}{r} \left(\lambda^2+\mu_{*}^2 r^2\right)^{\frac{1}{2}} R_2 \nonumber\\
&\exp\left[ -i \tan^{-1} \left(\dfrac{\mu_* r}{\lambda}\right)\right]. \label{radial8}
\end{align}

By substituting
\begin{align}
&R_1=\Phi_1 \exp\left[ -\dfrac{1}{2} i \tan^{-1} \left(\dfrac{\mu_* r}{\lambda}\right)\right],\\
&R_2=\Phi_2 \exp\left[\dfrac{1}{2} i \tan^{-1} \left(\dfrac{\mu_* r}{\lambda}\right)\right],
\end{align}
in Eqs. \eqref{radial7} and \eqref{radial8}, one can obtain
\begin{align}
&\dfrac{d \Phi_2}{dr_*}+i \omega \left[1+ \dfrac{f \mu_* \lambda}{2 \omega \sqrt{1+L} \left(\lambda^2+r^2 \mu_{*}^2 \right)} \right] \Phi_2 \nonumber\\
& =\dfrac{\sqrt{f}}{r} \left(\lambda^2+r^2 \mu_{*}^2 \right)^{\frac{1}{2}} \Phi_1, \label{radial9}\\
& \dfrac{d \Phi_1}{dr_*}-i \omega \left[1+ \dfrac{f \mu_* \lambda}{2 \omega \sqrt{1+L} \left(\lambda^2+r^2 \mu_{*}^2 \right)} \right] \Phi_1 \nonumber\\
& =\dfrac{\sqrt{f}}{r} \left(\lambda^2+r^2 \mu_{*}^2 \right)^{\frac{1}{2}} \Phi_2. \label{radial10}
\end{align}

Changing the variable $r_*$ into $\hat{r}_*$ as
\begin{align}\label{r*}
\hat{r}_*=r_*+\dfrac{1}{2 \omega} \tan^{-1} \left(\dfrac{\mu_* r}{\lambda}\right),
\end{align}
we get
\begin{align}
d\hat{r}_*= \left( \dfrac{\sqrt{1+L} \left(\lambda^2+r^2 \mu_{*}^2 \right)+\frac{f \mu_* \lambda}{2\omega}}{\sqrt{1+L} \left(\lambda^2+r^2 \mu_{*}^2 \right)}\right) d r_*.
\end{align}
Using Eq. \eqref{r*}, Eqs. \eqref{radial9} and \eqref{radial10} are simplified as
\begin{align}
\left( \dfrac{d}{d \hat{r}_*} +i \omega \right) \Phi_2= W \Phi_1, \label{radial11}\\
\left( \dfrac{d}{d \hat{r}_*} -i \omega \right) \Phi_1= W \Phi_2, \label{radial12}
\end{align}
where 
\begin{align}\label{W}
W=\dfrac{\sqrt{f}}{r} \dfrac{\sqrt{1+L} \left(\lambda^2+r^2 \mu_{*}^2 \right)^{\frac{3}{2}}}{\sqrt{1+L} \left(\lambda^2+r^2 \mu_{*}^2 \right)+\frac{f \mu_* \lambda}{2\omega}}.
\end{align}
Eqs. \eqref{radial11} and \eqref{radial12} can be combined  by letting
\begin{align}
\widetilde{Z}_+=\Phi_1+\Phi_2,\\
\widetilde{Z}_-=\Phi_1-\Phi_2.
\end{align} 
Thus, we  obtain the pair of one dimensional Schr\"{o}dinger-like wave equations
 
\begin{align}
\left(\dfrac{d^2}{d \hat{r}_{*}^{2} }+ \omega^2 \right) \widetilde{Z}_+=\widetilde{V_d}_+ \widetilde{Z}_+ \label{wavedm}\\
\left(\dfrac{d^2}{d \hat{r}_{*}^{2} }+ \omega^2 \right) \widetilde{Z}_-=\widetilde{V_d}_-\widetilde{Z}_-,
\end{align}
where $\widetilde{V_d}_\pm$ are the effective potentials, given by
\begin{align}\label{vmassivedirac}
\widetilde{V_d}_\pm=&W^2 \pm \dfrac{d W}{d \hat{r}_{*}}\nonumber\\
=&\dfrac{\widetilde{\Delta}^{\frac{1}{2}}\sqrt{1+L} ~\Xi^{\frac{3}{2}}}{\left[ r^2  \sqrt{1+L}~ \Xi+\frac{\Delta \mu_* \lambda}{2\omega} \right]^2}  \Biggl[ \widetilde{\Delta}^{\frac{1}{2}}\sqrt{1+L} ~\Xi^{\frac{3}{2}}  \nonumber\\
&\pm \biggl\{ 3 \widetilde{\Delta} r \mu_*^{2}+  \Xi \left(r-M-\dfrac{2}{3} \sqrt{1+L} \Lambda r^3 \right) \biggr\} \Biggr] \nonumber\\
&  \mp		\dfrac{\widetilde{\Delta}^{\frac{3}{2}}   \sqrt{1+L} ~\Xi^{\frac{5}{2}}}{\left[ r^2  \sqrt{1+L}~ \Xi+\frac{\widetilde{\Delta} \mu_* \lambda}{2\omega} \right]^3} \Biggl[ 2 \sqrt{1+L}~ r^3 \mu_{*}^2  \nonumber\\
&+2 r \sqrt{1+L} ~\Xi + \left(r-M-\dfrac{2}{3} \sqrt{1+L} \Lambda r^3 \right) \dfrac{\mu_* \lambda}{\omega} \Biggr],
\end{align}
where $\widetilde{\Delta}=r^2 f$ and $\Xi=\left(\lambda^2+r^2 \mu_{*}^2 \right)$.

The effective potential of massive Dirac field depends on $M$, $\Lambda$, $L$, $l$, $\mu$ and $\omega$. To study the effects of $l$, $\Lambda$ and $L$ on $\widetilde{V_d}_\pm$, we illustrate the effective potentials for different values of $l$, $\Lambda$ and $L$ in Fig \ref{fig vdm}. Similar to the effective potentials of massless Dirac field, the height of the peak point of effective potentials of massive Dirac field increase  with increasing $l$ and the position of the peak moves along the left for both  $\widetilde{V_d}_\pm$. However, with increasing $\Lambda$ and $L$, the height of the peak of effective potential decreases and the position of the peak moves toward the left. 
On comparing the positions of the peak of $\widetilde{V_d}_+$ and $\widetilde{V_d}_-$, it is observed that the peak of $\widetilde{V_d}_+$ occurs at  lower  values of $r$.
It is noted that  the overall nature of effective potentials of the scalar and Dirac field are the same. The effective potentials are positive definite between the event and cosmological horizons. They also have a single
maxima, which increases its height with increasing $l$  but they decrease  with increasing $\Lambda$ and $L$. The decrease  of the peak of effective potential with increasing Lorentz-violation parameter $L$ in the Bumblebee gravity model is also found in  \cite{uniyal2023,oliveira2021,lin2023,lambiase2023,jha2024}. Similarly, the effective potential decreases with increasing the loop quantum gravity parameter  and mass parameters (potential parameters) in loop quantum gravity theory \cite{jha2023b} and dRGT massive gravity theory \cite{boonserm2018,boonserm2023x} respectively. The effective potentials of the scalar and Dirac field will be used to analyze the behavior of the greybody factor in the next section.

\section{Greybody radiation}\label{section greybody}

Black holes emit radiation called Hawking radiation due to quantum effects near the event horizon of the black holes. When the Hawking radiation propagates out from the event horizon, it interacts with the curved spacetime around the black hole and this interaction modifies the spectrum and intensity of the Hawking radiation that escapes to infinity. Therefore, an observer at an infinite distance observes the modified form of Hawking radiation. The quantity which can measure how much modified spectrum deviates from the black body spectrum is usually known as greybody factor.
 In this section, we will investigate the greybody factor of scalar and Dirac particles emitted from Schwarzschild-de Sitter-like black hole using the rigorous bound. The rigorous bound on the greybody factor  is given by \cite{visser1999,boonserm2008}

\begin{align}\label{grey1}
T \geq \sech^2  \left( \int_{-\infty}^{+\infty} \wp dr_* \right),
\end{align}
where 
\begin{align}\label{grey2}
\wp=\dfrac{1}{2 h(r_*)} \sqrt{[h'(r_*)]^2 + \left(\omega^2 -V-h^2 (r_*) \right)^2        }.
\end{align}
Here $h(r_*)$ is a positive function and it must hold the  conditions  $h(+\infty)=h(-\infty)=\omega$ \cite{visser1999}. One can simply set $h=\omega$. Thus, Eq. \eqref{grey1} reduces to
\begin{align}\label{grey3}
T \geq \sech^2 \left( \int_{-\infty}^{+\infty} \dfrac{V}{2\omega} dr_*\right).
\end{align}
Note that the tortoise coordinate $r_*$  can also be written as
\begin{align}\label{tortoisemassless}
r_*= &\dfrac{3}{\sqrt{1+L} \Lambda} \Biggl[ \dfrac{r_h \log(r-r_h)}{(r_h-r_-) (r_c-r_h)}  \nonumber\\& - \dfrac{r_c \log(r-r_c)}{(r_c-r_-)(r_c-r_h)}   -\dfrac{r_-  \log(r-r_-)}{(r_c-r_1) (r_h-r_1)}  \Biggr]  .
\end{align}
From Eq. \eqref{tortoisemassless}, one can see that if $r \rightarrow r_h$ and $r\rightarrow r_c$ then $r_* \rightarrow -\infty$ and $r_*\rightarrow +\infty$ respectively. Therefore Eq. \eqref{grey3} can be written as
\begin{align}\label{grey4}
T \geq \sech^2 \left( \int_{r_h}^{r_c} \dfrac{V}{2\omega} dr\right).
\end{align}

\subsection{Greybody factor of scalar particle}

In this section, we will find the  rigorous bound on the greybody factor of scalar particle. Using the effective potential \eqref{klein10} in Eq. \eqref{grey4}, we obtain
\begin{align}\label{greyb1}
T_{s} \geq \sech^2 \left(	 \dfrac{\sqrt{1+L}}{2\omega}	 \int_{r_h}^{r_c}  \left[\dfrac{f'}{r(1+L)}-\dfrac{\lambda_s}{r^2}+m^2	\right] dr\right).
\end{align}
The rigorous bound of the Schwarzschild-de Sitter-like black hole for massive scalar fields is obtained as
\begin{align}\label{greyb2}
T_s \geq \sech^2 \Bigg( & \dfrac{\sqrt{1+L}}{2\omega} \Bigg[ \left(m^2- \dfrac{2 \Lambda}{3}	\right) \left(r_c-r_h\right)	+\lambda_s	\nonumber\\   	 	 & \left(\dfrac{1}{r_c}-\dfrac{1}{r_h} \right)	 -\dfrac{M}{(1+L)} \left(\dfrac{1}{r_c^{2}}-\dfrac{1}{r_{h}^2} \right)	\Bigg] \Bigg).
\end{align}

\begin{center}
\begin{figure*}[]
 \centering
  \begin{subfigure}[b]{0.32\textwidth}
    \centering
    \includegraphics[height=150pt,width=120pt]{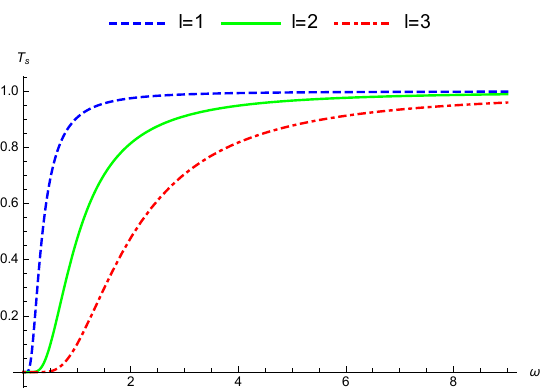} 
    \caption{}
    \label{fig Gsl}
  \end{subfigure}
  \hfill
  \begin{subfigure}[b]{0.32\textwidth}
    \centering
    \includegraphics[height=150pt,width=160pt]{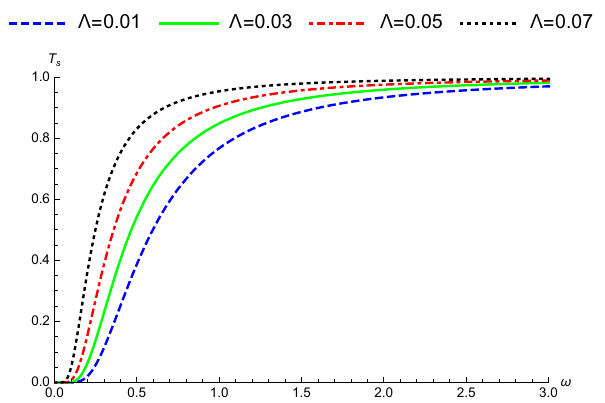} 
    \caption{}
    \label{fig GsLambda}
  \end{subfigure}
   \hfill
  \begin{subfigure}[b]{0.32\textwidth}
    \centering
    \includegraphics[height=150pt,width=160pt]{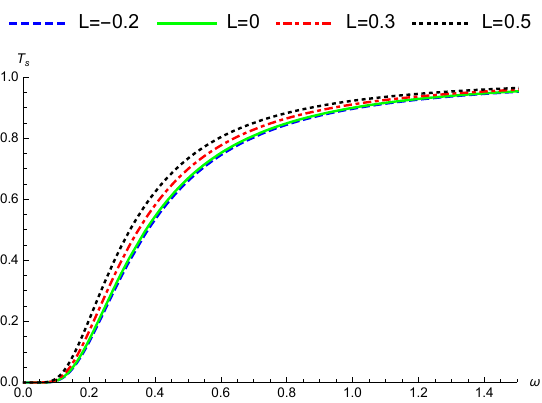} 
    \caption{}
    \label{fig GsLL}
    \end{subfigure}
  \caption{Variation of the Greybody factor for the massless scalar field: (a) for different values of $l$ with fixed $M=1$, $\Lambda=0.05$ and $L=0.2$; (b) for different values of $\Lambda$ with fixed $M=1$, $l=1$ and $L=0.2$; (c) for different values of $L$ with fixed $M=1$, $\Lambda=0.05$ and $l=1$. }
  \label{fig gs}
\end{figure*}
\end{center}

\begin{center}
\begin{figure*}[]
 \centering
  \begin{subfigure}[b]{0.32\textwidth}
    \centering
    \includegraphics[height=150pt,width=120pt]{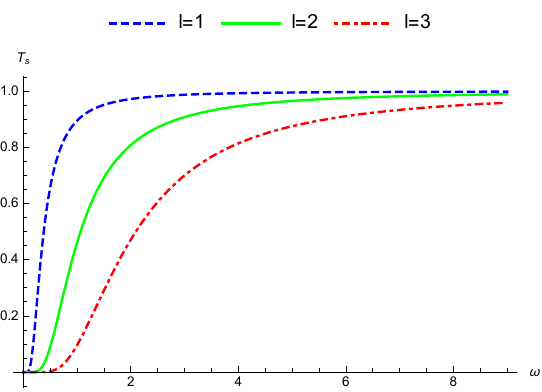} 
    \caption{}
    \label{fig Gsml}
  \end{subfigure}
  \hfill
  \begin{subfigure}[b]{0.32\textwidth}
    \centering
    \includegraphics[height=150pt,width=160pt]{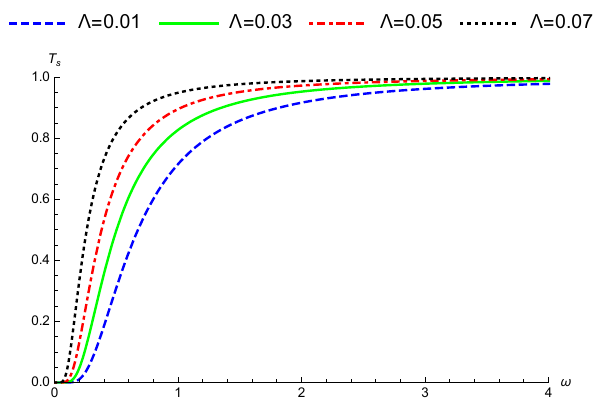} 
    \caption{}
    \label{fig GsmLambda}
  \end{subfigure}
   \hfill
  \begin{subfigure}[b]{0.32\textwidth}
    \centering
    \includegraphics[height=150pt,width=160pt]{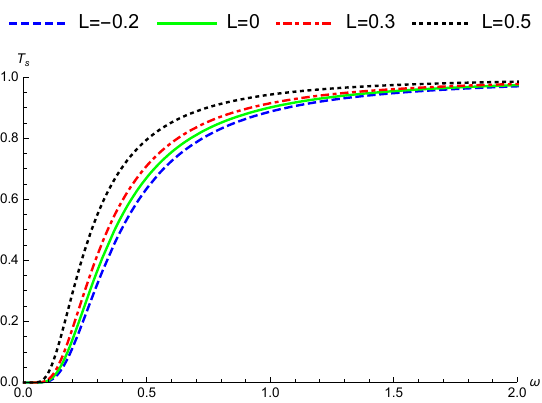} 
    \caption{}
    \label{fig GsmLL}
    \end{subfigure}
  \caption{Variation of  the rigorous bounds on the Greybody factor for the massive scalar field: (a) for different values of $l$ with fixed $M=1$, $\Lambda=0.05$ and $L=0.2$; (b) for different values of $\Lambda$ with fixed $M=1$, $l=1$, $m=0.1$ and $L=0.2$; (c) for different values of $L$ with fixed $M=1$, $\Lambda=0.05$,  $m=0.1$ and $l=1$. }
  \label{fig gsm}
\end{figure*}
\end{center}

\begin{center}
\begin{figure*}[]
 \centering
  \begin{subfigure}[b]{0.32\textwidth}
    \centering
    \includegraphics[height=150pt,width=150pt]{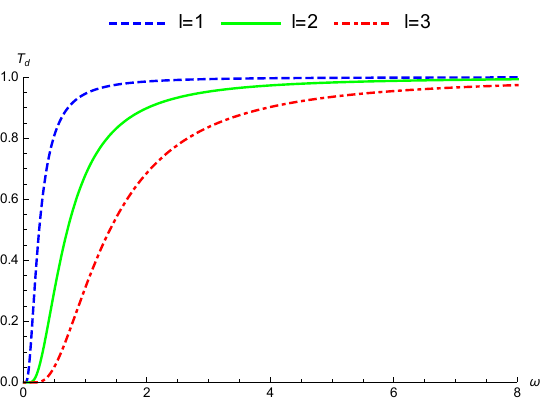} 
    \caption{}
    \label{fig gdl}
  \end{subfigure}
  \hfill
  \begin{subfigure}[b]{0.32\textwidth}
    \centering
    \includegraphics[height=150pt,width=150pt]{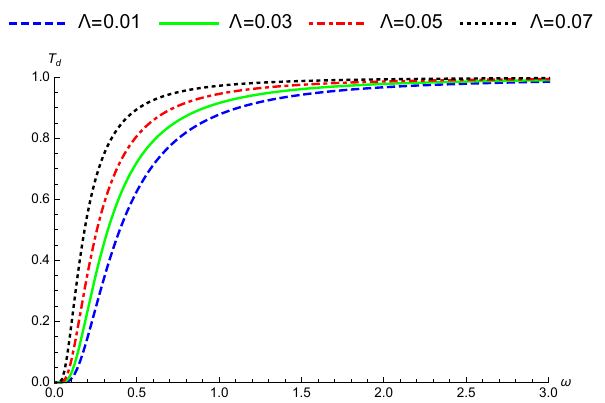} 
    \caption{}
    \label{fig gdL}
  \end{subfigure}
   \hfill
  \begin{subfigure}[b]{0.32\textwidth}
    \centering
    \includegraphics[height=150pt,width=150pt]{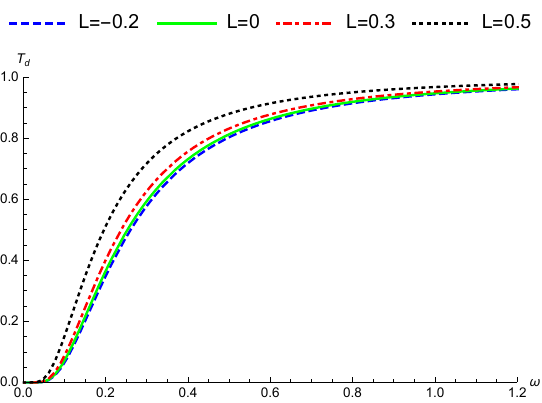} 
    \caption{}
    \label{fig gsLL}
    \end{subfigure}
  \caption{Variation of the rigorous bounds on the greybody factor for the massless Dirac field: (a) for different values of $l$ with fixed $M=1$, $\Lambda=0.05$ and $L=0.2$; (b) for different values of $\Lambda$ with fixed $M=1$, $l=1$ and $L=0.2$; (c) for different values of $L$ with fixed $M=1$, $\Lambda=0.05$ and $l=1$.}
  \label{fig gd}
\end{figure*}
\end{center}

If $m=0$, Eq. \eqref{greyb2} reduces to the rigorous bound of the Schwarzschild-de Sitter-like black hole for massless scalar fields. We depict the greybody factors of massless and massive scalar fields in Figs. \ref{fig gs} and \ref{fig gsm} respectively. We show the effects of $l$, $\Lambda$ and $L$  on the greybody factors.  For both cases, we observe that the greybody factors are suppressed by $l$ when the other parameters are fixed. This point is consistent with the earlier results found in Sec. \ref{section scalar}
that increasing  $l$ increases the potential barrier, so the wave is more likely to be reflected thereby resulting the low greybody factor. Further, the greybody factors are monotonically increasing with $\Lambda$ and $L$.  These larger greybody factors result from the fact that the parameters $\Lambda$ and $L$ suppressed the potential barrier, making the wave less probable to be reflected.

\subsection{Greybody factor of fermions}
\subsubsection{Massless fermions}
Here, we will derive the greybody factor of the neutrinos emitted from the  Schwarzschild-de Sitter-like black hole. Using the effective potential \eqref{vmasslessdirac} in Eq. \eqref{grey4}, one can derive the rigorous bound of the Schwarzschild-de Sitter-like black hole for neutrinos as
\begin{align}\label{greyfermionmassless}
T_d \geq  \sech^2 &	 \left[ \int_{r_h}^{r_c}  \dfrac{\sqrt{1+L}}{2\omega f}  \Biggl( \dfrac{\lambda_{f}^2 f}{r^2}\pm \dfrac{\lambda_f  \sqrt{f} \left(3M-r \right)}{r^3 \sqrt{1+L}} \Biggr) dr	\right] \nonumber\\
=\sech^2 & \Biggl[	\dfrac{1}{2 \omega} \Bigg\{ - \sqrt{1+L}~\lambda_f^{2} \left(\dfrac{1}{r_c}-\dfrac{1}{r_h}\right) \nonumber\\ &  \pm\lambda_f  \left(\dfrac{\sqrt{f(r_c)}}{r_c}-\dfrac{\sqrt{f(r_h)}}{r_h} \right) \Bigg\} \Biggr].
\end{align}

It is worth noting that $f(r_h)=f(r_c)=0$, therefore  Eq. \eqref{greyfermionmassless} reduces to
\begin{align}\label{greyfermionmassless2}
T_d \geq  
\sech^2 & \Biggl[	\dfrac{1}{2 \omega} \Bigg\{ - \sqrt{1+L}~\lambda_f^{2} \left(\dfrac{1}{r_c}-\dfrac{1}{r_h}\right) \Biggr].
\end{align}


\subsubsection{Massive fermions}

From Eq. \eqref{grey3}, the rigorous bound of the Schwarzschild-de Sitter-like black hole for the massive fermions is obtained as
\begin{align} \label{greymassive1}
\widetilde{T}_{d}  \geq  \sech^2 \Biggl(  \int_{-\infty}^{+\infty}  \dfrac{\widetilde{V}_\pm }{2\omega} d\hat{r}_{*} \Biggr) 
\end{align}
Note that $\hat{r}_*$  can be written as
\begin{align}\label{tortoisemassive}
\hat{r}_*= &\dfrac{3}{\sqrt{1+L} \Lambda} \Biggl[ \dfrac{r_h \log(r-r_h)}{(r_h-r_-) (r_c-r_h)}  \nonumber\\& - \dfrac{r_c \log(r-r_c)}{(r_c-r_-)(r_c-r_h)}   -\dfrac{r_-  \log(r-r_-)}{(r_c-r_1) (r_h-r_1)}  \Biggr]  \nonumber\\
&+ \dfrac{1}{2\omega} \tan^-1 \left(\dfrac{\mu_* r}{\lambda_f} \right).
\end{align}
One can see from Eq. \eqref{tortoisemassive} that $\hat{r}_{*} \rightarrow -\infty$ when $r\rightarrow r_h$ and  $\hat{r}_{*} \rightarrow +\infty$ when $r\rightarrow r_c$. Thus, Eq. \eqref{greymassive1} reduces to
\begin{align}\label{greymassive3}
\widetilde{T}_{d}  \geq  \sech^2  \Biggl[ & \dfrac{1}{2\omega}  \Biggl\{ \int_{r_h}^{r_c}  \dfrac{(1+L) \Xi^2}{r^2 \sqrt{1+L} \Xi + \dfrac{\Delta \mu_* \lambda_f}{2\omega}}   dr  \nonumber\\ & \pm W|_{r_h}^{r_c} \Biggr\}\Biggr].
\end{align}

We observe from Eq. \eqref{W} that  $W$ is proportional to $\sqrt{f}$ and since $f$ vanishes at the event and cosmological horizons, it is worth noting that $W|_{r_h}^{r_c}=0$. Further,  the integral part in Eq. \eqref{greymassive3} can be solved analytically, however it is difficult to investigate the  behaviour of the greybody factor. Therefore, we use the approximation technique to investigate the greybody factor bound.   The integral can be expressed in the following form
\begin{align}\label{greymassive4}
\int_{r_h}^{r_c} \dfrac{ \lambda_f^{2} \sqrt{1+L} ~ (1+r^2 \tilde{\mu}^2)}{r^2 B}  dr,
\end{align}
where $\tilde{\mu}=\dfrac{\mu_*}{\lambda_f}$  and $B=1+ \dfrac{f \tilde{\mu}}{\sqrt{1+L}(1+r^2 \tilde{\mu}^2) 2 \omega } $.
In the region $r_h<r<r_c$, $f$ is always positive, thus the quantity $B>1$.
 So one can  use the inequality $A\leq A_{approx}$, where
\begin{align}
&A= \dfrac{ \lambda_f^{2} \sqrt{1+L} ~ (1+r^2 \tilde{\mu}^2)}{r^2 B}, \nonumber\\
& A_{approx}=  \dfrac{\lambda_l^{2}  \sqrt{1+L} (1+r^2 \tilde{\mu}^2) }{r^2}.
\end{align}
It is worth mentioning that both the original and approximated integrand  of Eq. \eqref{greymassive4} are positive, thus one can approximate  Eq. \eqref{greymassive4} as

\begin{align}
\int_{r_h}^{r_c} A dr \leq \int_{r_h}^{r_c} A_{approx} dr .
\end{align}
It follows that
\begin{align}\label{greymassive7}
\widetilde{T}_{d}  \geq & \sech^2  \biggl[  \dfrac{1}{2\omega}   \int_{r_h}^{r_c}  A  ~dr    \biggr]  \nonumber\\
 \geq &\sech^2  \biggl[  \dfrac{1}{2\omega}  \int_{r_h}^{r_c}  A_{approx}  ~dr    \biggr].
\end{align}
Now the integral of Eq. \eqref{greymassive7} is calculated as
\begin{align}
\int_{r_h}^{r_c} A_{approx} dr= \sqrt{1+L} \left[ \mu_{*}^2 \left(r_c-r_h\right) -\lambda_{f}^2 \left(\dfrac{1}{r_c^{2}}-\dfrac{1}{r_h^{2}} \right) \right].
\end{align}
The rigorous bound of the Schwarzschild-de Sitter-like black hole for massive fermions is found to be
\begin{align}
\widetilde{T}_{d}  \geq  \sech\Biggl[ \dfrac{\sqrt{1+L}}{2\omega} \left\lbrace \mu_{*}^2 \left(r_c-r_h\right) -\lambda_{f}^2 \left(\dfrac{1}{r_c^{2}}-\dfrac{1}{r_h^{2}} \right) \right\rbrace \Biggr].
\end{align}

%
%
\begin{center}
\begin{figure*}[]
 \centering
  \begin{subfigure}[b]{0.32\textwidth}
    \centering
    \includegraphics[height=150pt,width=150pt]{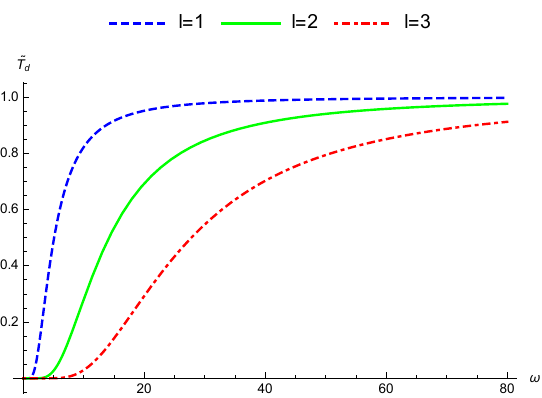} 
    \caption{}
    \label{fig gdml}
  \end{subfigure}
  \hfill
  \begin{subfigure}[b]{0.32\textwidth}
    \centering
    \includegraphics[height=150pt,width=150pt]{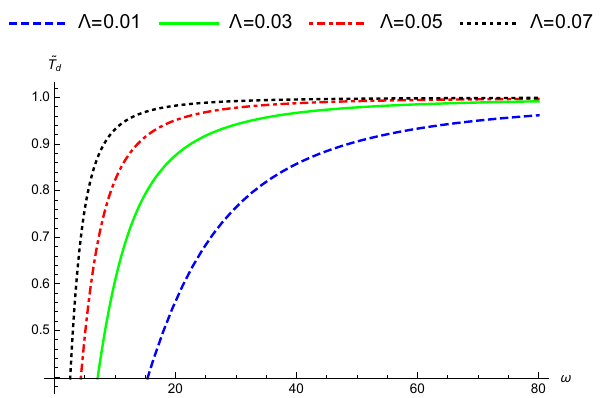} 
    \caption{}
    \label{fig gdmL}
  \end{subfigure}
   \hfill
  \begin{subfigure}[b]{0.32\textwidth}
    \centering
    \includegraphics[height=150pt,width=150pt]{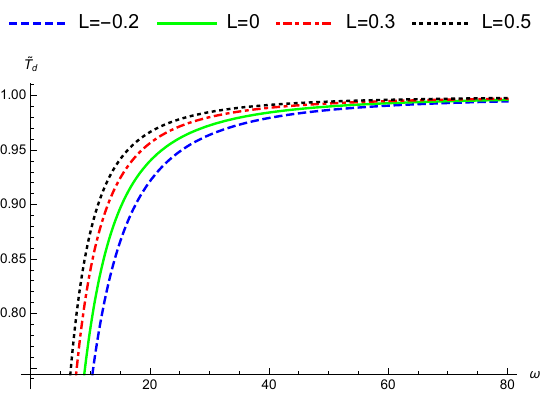} 
    \caption{}
    \label{fig gdmLL}
    \end{subfigure}
  \caption{Variation of the rigorous bounds on the greybody factor for the massive Dirac field: (a) for different values of $l$ with fixed $M=1$, $\Lambda=0.05$, $\mu=1$ and $L=0.2$; (b) for different values of $\Lambda$ with fixed $M=1$, $l=1$, $\mu=1$ and $L=0.2$; (c) for different values of $L$ with fixed $M=1$, $\Lambda=0.05$, $\mu=1$ and $l=1$.  }
  \label{fig gdm}
\end{figure*}
\end{center}

The variation of the rigorous bound on the greybody factors of massless and massive Dirac field are illustrated in Figs. \ref{fig gd} and \ref{fig gdm} respectively. The behaviour of the rigorous bound on the greybody factors of Dirac field show a similar behaviour to that of scalar field.
The greybody factors decreases with increasing $l$ but it increases with increasing $\Lambda$ and $L$.  We observe that the rigorous bounds on the greybody factors of massive case have larger variation in comparison to the massless case. 


\section{Quasinormal mode of Schwarzschild-de Sitter-like black hole}\label{section quasi}

\begin{figure*}[]
 \centering
  \begin{subfigure}[b]{0.32\textwidth}
    \centering
    \includegraphics[width=\textwidth]{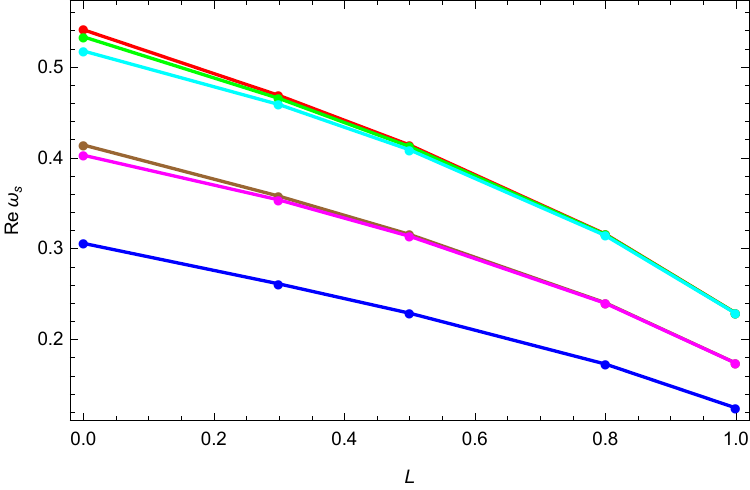} 
    \caption{Real part of scalar quasinormal frequencies for different values of $L$.}
  \end{subfigure}
  \hfill
  \begin{subfigure}[b]{0.32\textwidth}
    \centering
    \includegraphics[width=\textwidth]{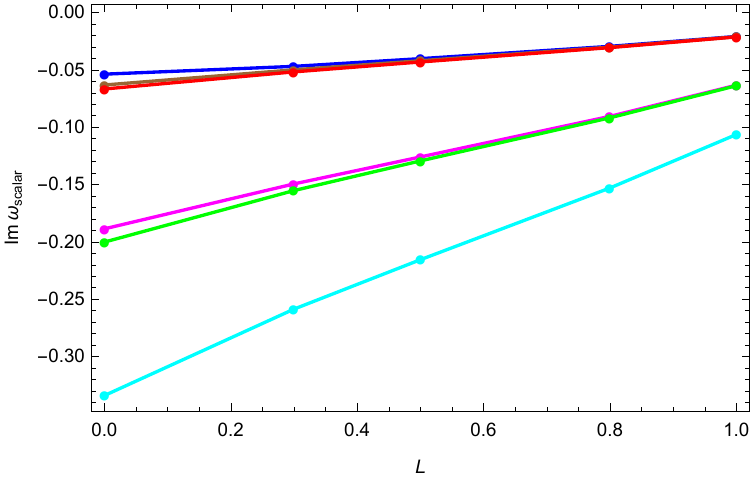} 
    \caption{Imaginary part of scalar quasinormal frequencies for different values of $L$.}
  \end{subfigure}
   \hfill
  \begin{subfigure}[b]{0.32\textwidth}
    \centering
    \includegraphics[width=\textwidth]{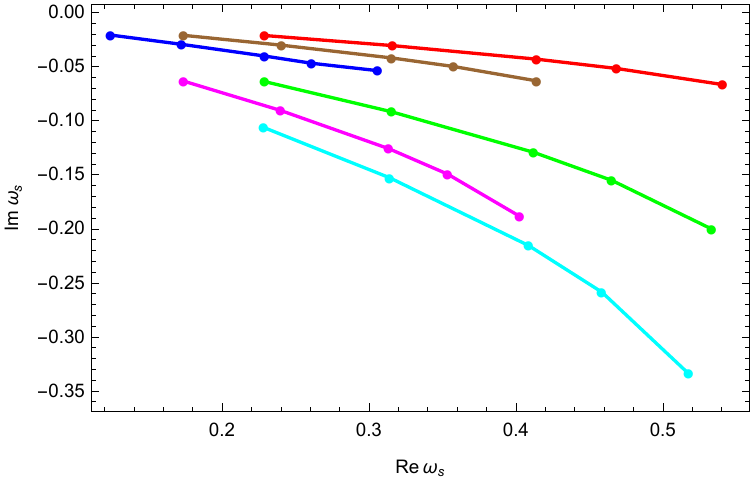} 
    \caption{Scalar quasinormal frequencies varying $L$.}
    \end{subfigure}
    \hfill
  \begin{subfigure}[b]{0.32\textwidth}
    \centering
    \includegraphics[width=\textwidth]{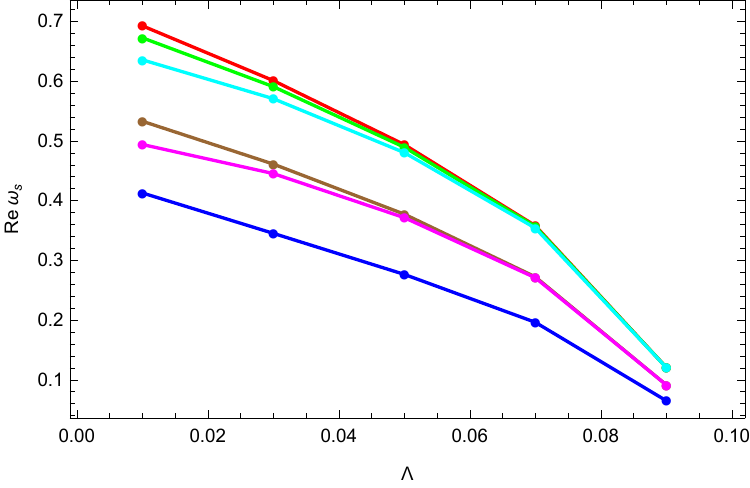} 
    \caption{Real part of scalar quasinormal frequencies for different values of $\Lambda$.}
  \end{subfigure}
  \hfill
  \begin{subfigure}[b]{0.32\textwidth}
    \centering
    \includegraphics[width=\textwidth]{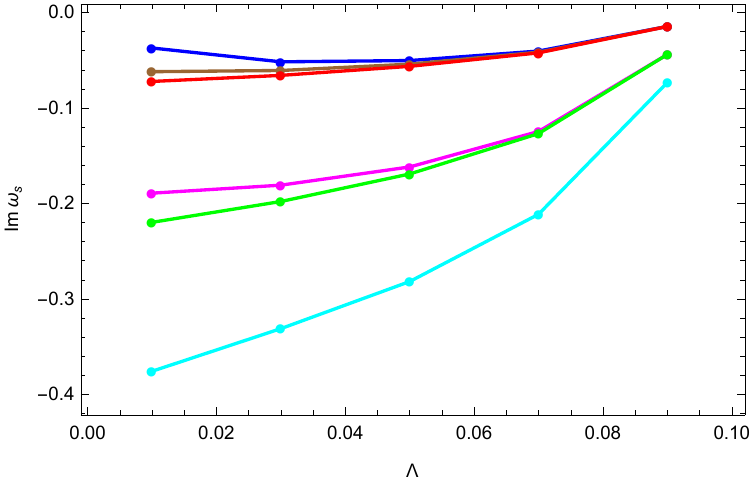} 
    \caption{Imaginary part of scalar quasinormal frequencies for different values of $\Lambda$.}
  \end{subfigure}
  \hfill
  \begin{subfigure}[b]{0.32\textwidth}
    \centering
    \includegraphics[width=\textwidth]{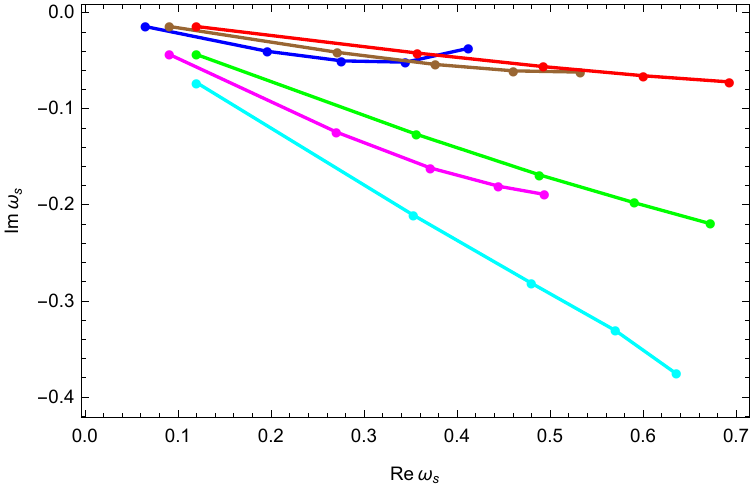} 
    \caption{Scalar quasinormal frequencies varying $\Lambda$.}
  \end{subfigure}
  \caption{Scalar quasinormal frequencies for a fixed $M=1$ and $m=0.5$. For Figs. (a)-(c), we set $\Lambda=0.05$  and  for Figs. (d)-(f), we set $L=0.2$. The colors corresponding to the different values of $(l,n)$ are blue (1,0); brown (2,0); magenta (2,1); red(3,0); green (3,1) and cyan (3,2).}
  \label{qnms}
\end{figure*}


\begin{center}
\begin{figure*}[h!]
 \centering
  \begin{subfigure}[b]{0.32\textwidth}
    \centering
    \includegraphics[width=\textwidth]{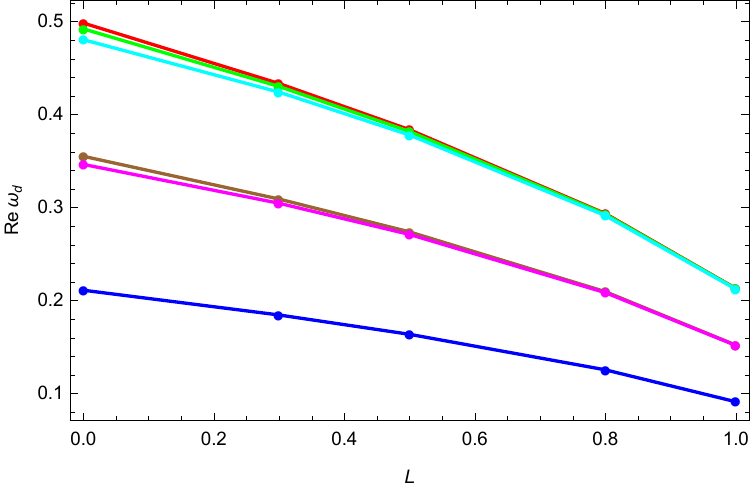} 
    \caption{Real part of massless Dirac quasinormal frequencies for different values of $L$.}
    \label{fig:figure1}
  \end{subfigure}
  \hfill
  \begin{subfigure}[b]{0.32\textwidth}
    \centering
    \includegraphics[width=\textwidth]{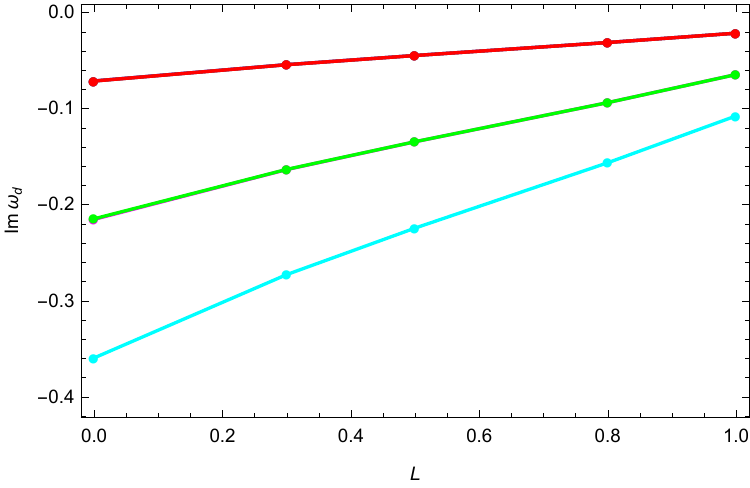} 
    \caption{Imaginary part of massless Dirac quasinormal frequencies for different values of $L$.}
    \label{fig:figure2}
  \end{subfigure}
   \hfill
  \begin{subfigure}[b]{0.32\textwidth}
    \centering
    \includegraphics[width=\textwidth]{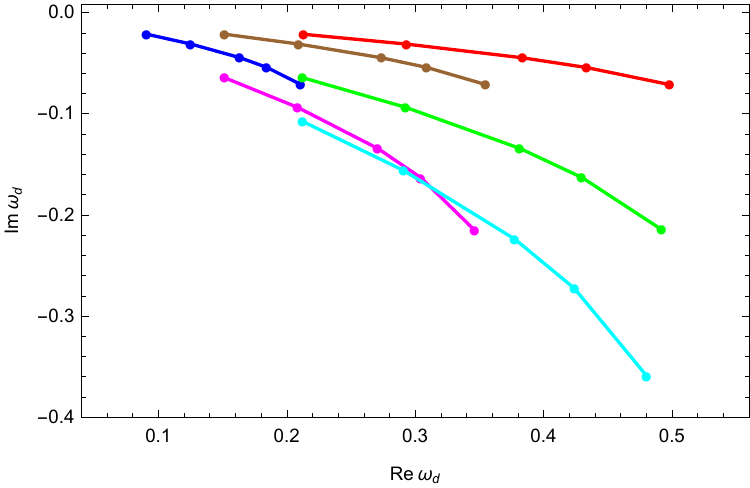} 
    \caption{Massless Dirac quasinormal frequencies varying $L$.}
    \label{fig:figure5}
    \end{subfigure}
    \hfill
  \begin{subfigure}[b]{0.32\textwidth}
    \centering
    \includegraphics[width=\textwidth]{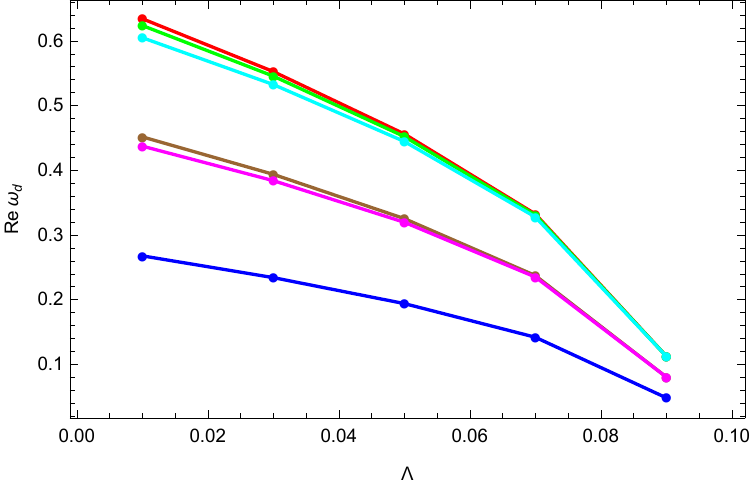} 
    \caption{Real part of massless Dirac quasinormal frequencies for different values of $\Lambda$.}
    \label{fig:figure3}
  \end{subfigure}
  \hfill
  \begin{subfigure}[b]{0.32\textwidth}
    \centering
    \includegraphics[width=\textwidth]{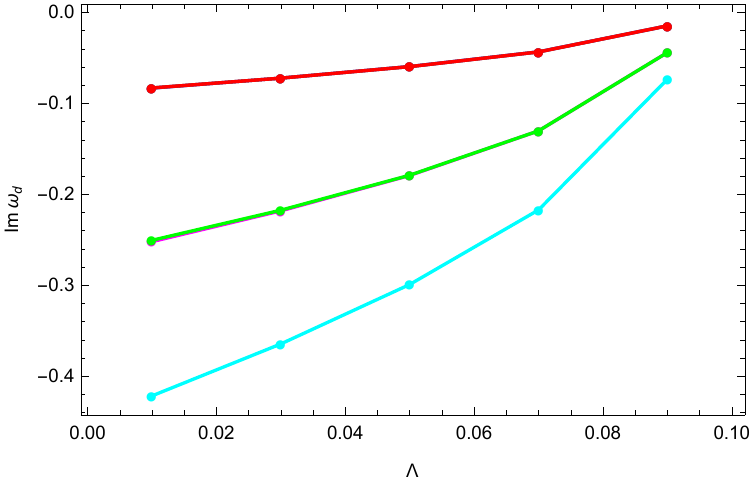} 
    \caption{Imaginary part of massless Dirac quasinormal frequencies for different values of $\Lambda$.}
    \label{fig:figure4}
  \end{subfigure}
  \hfill
  \begin{subfigure}[b]{0.32\textwidth}
    \centering
    \includegraphics[width=\textwidth]{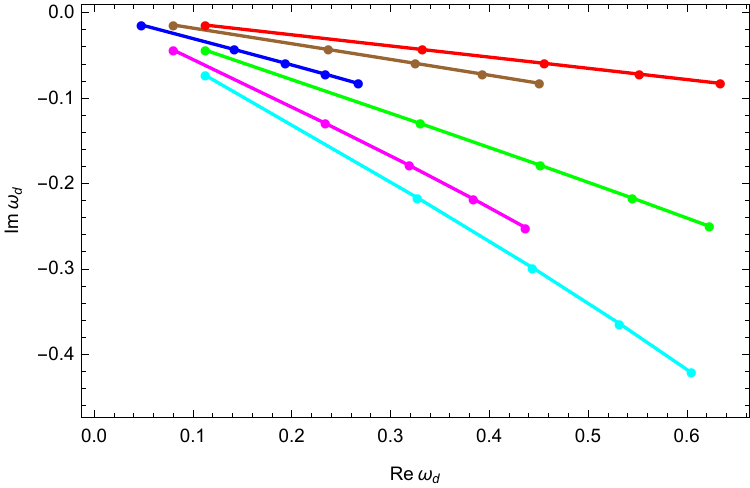} 
    \caption{Massless Dirac quasinormal frequencies varying $\Lambda$.}
    \label{fig:figure6}
  \end{subfigure}
\caption{Massless Dirac quasinormal frequencies for a fixed $M=1$. For Figs. (a)-(c), we set $\Lambda=0.05$  and for Figs. (d)-(f), we set $L=0.2$. The colors corresponding to the different values of $(l,n)$ are blue (1,0); brown (2,0); magenta (2,1); red(3,0); green (3,1) and cyan (3,2).}
\label{qnmmassless}
\end{figure*}
\end{center}


\begin{center}
\begin{figure*}[]
 \centering
  \begin{subfigure}[b]{0.32\textwidth}
    \centering
    \includegraphics[width=\textwidth]{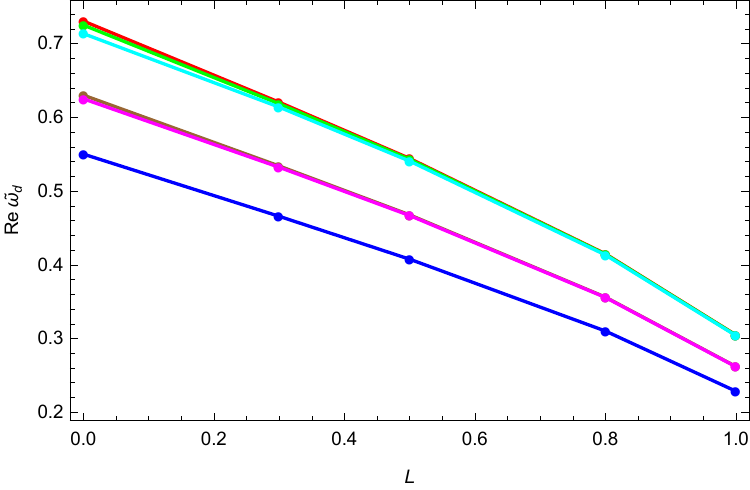} 
    \caption{Real part of massive Dirac quasinormal frequencies for different values of $L$.}
    \label{fig:figure1}
  \end{subfigure}
  \hfill
  \begin{subfigure}[b]{0.32\textwidth}
    \centering
    \includegraphics[width=\textwidth]{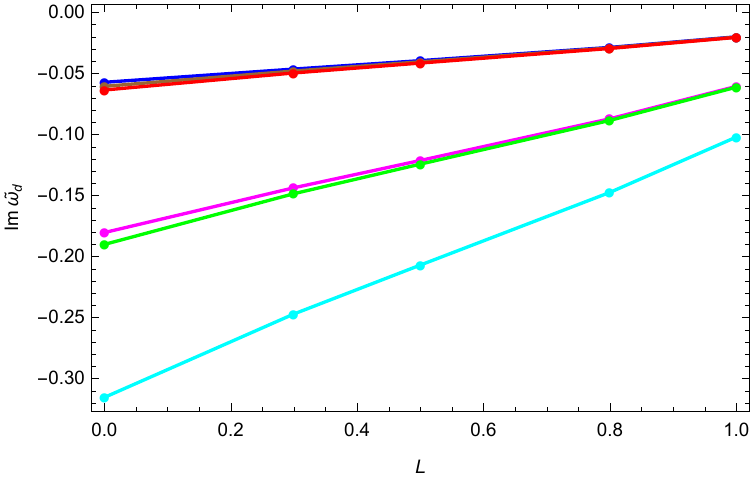} 
    \caption{Imaginary part of  massive Dirac quasinormal frequencies for different values of $L$.}
    \label{fig:figure2}
  \end{subfigure}
   \hfill
  \begin{subfigure}[b]{0.32\textwidth}
    \centering
    \includegraphics[width=\textwidth]{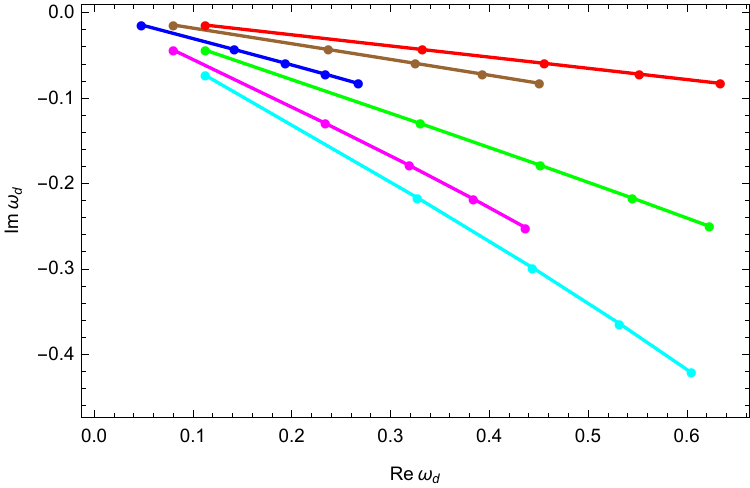} 
    \caption{Massive Dirac quasinormal frequencies varying $L$.}
    \label{fig:figure5}
    \end{subfigure}
    \hfill
  \begin{subfigure}[b]{0.32\textwidth}
    \centering
    \includegraphics[width=\textwidth]{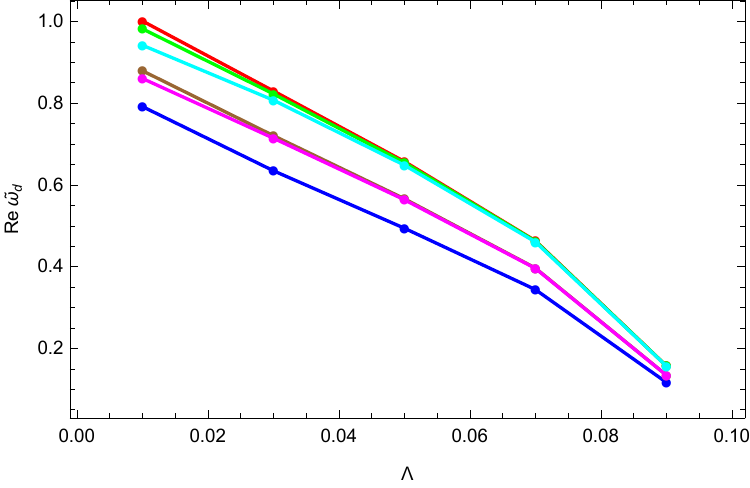} 
    \caption{Real part of  massive Dirac quasinormal frequencies for different values of $\Lambda$.}
    \label{fig:figure3}
  \end{subfigure}
  \hfill
  \begin{subfigure}[b]{0.32\textwidth}
    \centering
    \includegraphics[width=\textwidth]{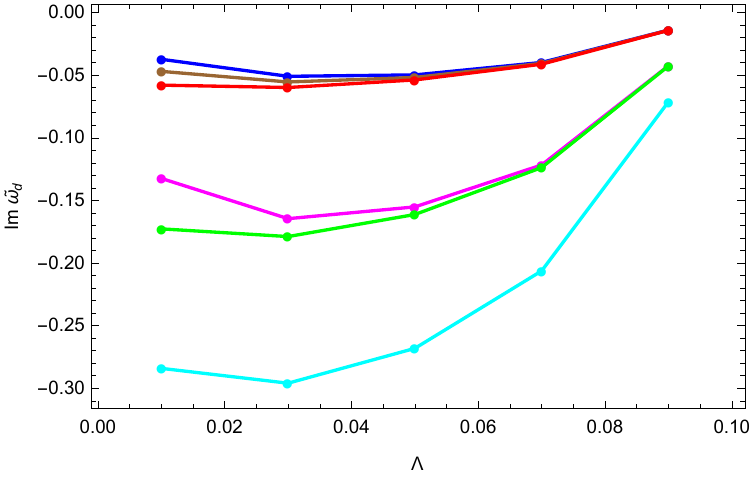} 
    \caption{Imaginary part of  massive Dirac quasinormal frequencies for different values of $\Lambda$.}
    \label{fig:figure4}
  \end{subfigure}
  \hfill
  \begin{subfigure}[b]{0.32\textwidth}
    \centering
    \includegraphics[width=\textwidth]{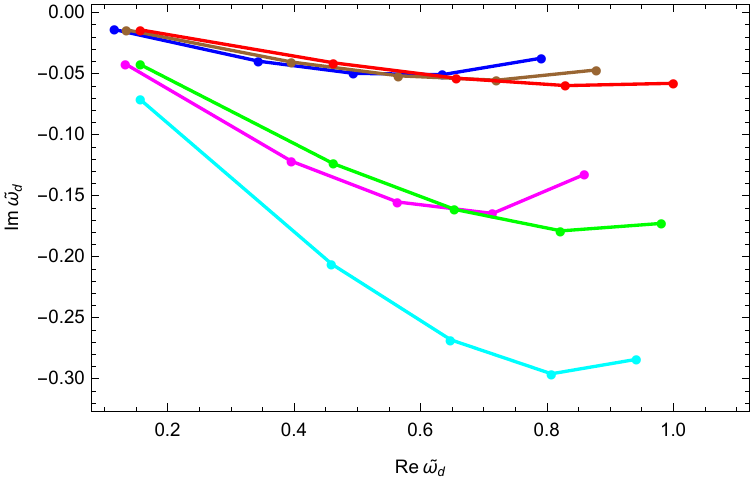} 
    \caption{Massive Dirac quasinormal frequencies varying $\Lambda$.}
    \label{fig:figure6}
  \end{subfigure}
  \label{fig qnms}\caption{Massive Dirac quasinormal frequencies for a fixed $M=1$,  $\omega=0.2$ and $\mu_*=1$.  For Figs. (a)-(c), we set $\Lambda=0.05$ and for Figs. (d)-(f), we set $L=0.2$ . The colors corresponding to the different values of $(l,n)$ are blue (1,0); brown (2,0); magenta (2,1); red(3,0); green (3,1) and cyan (3,2).}
\end{figure*}
\end{center}

In this section, we will consider the scalar and Dirac fields perturbations of the Schwarzschild-de Sitter-like black hole to investigate the behaviour of quasinormal modes. The quasinormal modes of scalar and Dirac fields perturbations are the solutions of wave equations \eqref{klein9}, \eqref{waved} and \eqref{wavedm} which satisfy specific boundary conditions at the black hole horizon and far from the black hole \cite{konoplya2011}. The solution must satisfy the conditions of purely ingoing waves at the event horizon 
and purely outgoing waves at  cosmological horizon or spatial infinity.
\subsection{Quasinormal modes via WKB method}
We  will study the quasinormal modes of scalar and Dirac fields using the semi-analytical methods based on WKB approximation. Schutz and Will \cite{schutz1985} proposed the WKB method for determining the quasinormal of black hole. Later Iyer and Will \cite{iyer1987b,iyer1987a} extended upto third-order. The formula for finding the quasinormal frequencies using the third order WKB method is given by 

\begin{align}\label{q1}
\omega^2=&\biggl[ V_0+\left(-2 V_{0}^{''}\right)^{\frac{1}{2}} \tilde{\Lambda}(n)\nonumber\\ 
&  -i  \left(n+\frac{1}{2} \right) \left(-2 V_{0}^{''}\right)^{\frac{1}{2}} \left[ 1+\tilde{\Omega}(n) \right]	\biggr],
\end{align}
where 
\begin{align}
\Lambda(n)=& \dfrac{1}{\left(-2 V_{0}^{''}\right)^{1/2}} \Biggl[ \dfrac{1}{8} \left(\dfrac{V_{0}^{(4)}}{V_{0}^{''}}\right) \left(\dfrac{1}{4}+\alpha^2 \right)\nonumber\\
& - \dfrac{1}{288}  \left(\dfrac{V_{0}^{(3)}}{V_{0}^{''}}\right)^2  \left(7+60 \alpha^2 \right) \Biggr],
\end{align}
\begin{align}
 \Omega(n) = &\dfrac{1}{\left(-2 V_{0}^{''}\right)} \Biggl[ \dfrac{5}{6912} \left(\dfrac{V_{0}^{(3)}}{V_{0}^{''}}\right)^4 \left(77+188 \alpha^2 \right)		\nonumber\\
 & -\dfrac{1}{384} \left(\dfrac{V_{0}^{(3)^2} V_{0}^{(4)}}{V_{0}^{''^3}}\right)	\left(51+100 \alpha^2\right) +\dfrac{1}{2304} \nonumber\\
 & \times  \left(\dfrac{V_{0}^{(4)}}{V_{0}^{''}}\right)^2 \left(67+68 \alpha^2 \right) +\dfrac{1}{288} \left(\dfrac{V_{0}^{(3)} V_{0}^{(5)}}{V_{0}^{''^2}}\right) \nonumber\\
 & \times \left(19+28 \alpha^2\right) -\dfrac{1}{288} \left(\dfrac{V_{0}^{(6)}}{V_{0}^{''}}\right) \left(5+4 \alpha^2 \right) \Biggr],
\end{align}
where the primes and the superscript $(n)$ denote the differentiation of the effective potential with respect to the tortoise coordinate $r_*$.  $V_0$ is the  value of potential calculated at $r_0$, where $r_0$ is the location of the peak of $V$ and $\alpha= n+\frac{1}{2}$ where $n$ is a positive integer which denotes the overtone number. The multipole number $(l)$ and the overtone number $(n)$ are critical factors that determine the accuracy of the WKB method. For $l>n$, the accuracy of the 3rd order WKB approximation is excellent.

\subsection{Quasinormal modes via  P\"{o}sch–Teller fitting method}

In this method, the P\"{o}schl–Teller function \cite{ferrari1984} is used to approximate the effective potentials. The P\"{o}schl–Teller potential is given by
\begin{align}
V_{PT}=\dfrac{V_0}{\cosh^2 \alpha (r_*-r_{*_0})},
\end{align}
where $V_0$ and $\alpha$ represent the height and curvature of the effective potential at its maximum $(r_*=r_{*_0})$, respectively. The bound states of the Pöschl-Teller potential is given by
\begin{align}
\Omega_{n}(V_0,\alpha)=\alpha \left[	-\left(n+\dfrac{1}{2} \right)+\left(\dfrac{1}{4}+\dfrac{V_0}{\alpha^2}\right)^{1/2}		\right].
\end{align}
Using the inverse transformations, the quasinormal modes $\omega$ can be obtained as
\begin{align}\label{ptfm}
\omega=\pm \sqrt{V_0-\dfrac{\alpha^2}{4}}-i \alpha \left(n+\dfrac{1}{2}\right).
\end{align}


\subsection{Scalar perturbation}
Using the effective potential \ref{klein10}, we obtain the numerically calculated quasinormal frequencies for the scalar perturbation using the 3rd order WKB approximation. The quasinormal frequencies is also calculated by using the P\"{o}schl–Teller fitting method from Eq. \eqref{ptfm}. 
The quasinormal frequencies obtained from both the methods by varying the parameters $L$ and $\Lambda$ are displayed in Tables \ref{tab qnms1} and \ref{tab qnms2} respectively. The case of $L=0$ corresponds to the original Schwarzschild-de Sitter black hole. The quasinormal frequencies obtained by using WKB 3rd order are also summarized in Fig. \ref{qnms}.

\begin{table}[h!]
\caption{Quasinormal frequencies for  massive scalar perturbation $\omega_{scalar}$ for different modes and for different values of the Lorentz violating parameter $L$ with fixed value of $\Lambda=0.05$ and $m=0.5$. }
\label{tab qnms1}
\begin{tabular}{p{0.24cm} p{0.24cm} p{0.24cm} p{2.8cm} p{2.8cm}}
\hline
 $L$& $l$	  & $n$  &  WKB & P\"{o}schl–Teller fitting \\ \hline 
0 & 1 & 0 &   0.305834-0.0537772$i$ &	0.305935 - 0.057241$i$\\
 & 2 & 0 & 0.414064-0.063237$i$ &  0.415646 - 0.063904$i$\\
 &  & 1 & 0.402809-0.188693$i$ & 0.415646 - 0.191715$i$\\
  & 3 & 0 & 0.540911-0.0667219$i$ & 0.542196 - 0.067085$i$ \\
 &  &  1&0.533167-0.200246$i$ & 0.542196 - 0.201256$i$\\
 &  & 2 &   0.517556-0.334103$i$ & 0.542196 - 0.335427$i$\\
 \hline
 0.3 & 1 & 0 & 0.261165-0.0469813$i$ & 0.260948 - 0.046753$i$ \\
 & 2 & 0 &   0.357733-0.0499821$i$ & 0.358282 - 0.050179$i$ \\
 &  & 1 & 0.353582-0.149512$i$ & 0.358282 - 0.150538$i$ \\
  & 3 & 0 &  0.46857-0.0517846$i$ & 0.469099 - 0.051921$i$ \\
 &  &  1&  0.465281-0.155297$i$ & 0.469099 - 0.155765$i$\\
 &  & 2 & 0.45857-0.258709$i$ & 0.469099 - 0.259609$i$\\
 \hline
 0.5	&	1&	0& 0.228933-0.0402517$i$ & 0.228751 - 0.040084$i$	\\
 	&	2&	0& 0.315617-0.0420673$i$	 & 0.315871 - 0.042144$i$\\
 	&	&	1& 0.313593-0.126004$i$ &	0.315871 - 0.126433$i$\\
 & 3 & 0 &  0.414069-0.0431606$i$ & 0.414346 - 0.043226$i$\\
 &  &  1& 0.412314-0.12944$i$ & 0.414346 - 0.12968$i$\\
 &  & 2 & 0.408731-0.215631$i$ &  0.414346 - 0.216133$i$\\
 	\hline
0.8 & 1 & 0 & 0.172959-0.0295259$i$& 0.172875 - 0.029457$i$\\
 & 2 & 0 & 0.240382-0.0302083$i$ & 0.240443 - 0.030219$i$ \\
 &  & 1 & 0.239833-0.0905857$i$ & 0.240443 - 0.090659$i$\\
 & 3 & 0 &  0.316041-0.0306193$i$& 0.316124 - 0.030636$i$ \\
 &  &  1& 0.315496-0.0918468$i$& 0.316124 - 0.091908$i$ \\
 &  & 2 & 0.314389-0.153048$i$ & 0.316124 - 0.15318$i$\\
 \hline
1 	&	1&0	&   0.124626-0.0209119$i$	& 0.124594 - 0.020887$i$\\
 	&	2&0	&	0.173998-0.0211467$i$ & 0.174013 - 0.021147$i$\\
 	&	&1	&0.173847-0.0634335$i$	& 0.174013 - 0.063442$i$\\
 & 3 & 0 & 0.229051-0.0212884$i$ & 0.229076 - 0.021292$i$\\
 &  &  1& 0.228886-0.0638632$i$&  0.229076 - 0.063877$i$\\
 &  & 2 & 0.228553-0.106434$i$ & 0.229076 - 0.106463$i$\\
 	\hline
\end{tabular}
\end{table}

\begin{table}[h]
\caption{Quasinormal frequencies for  massive scalar perturbation for different modes and for different values of cosmological constant $\Lambda$ with fixed value of $L=0.2$ and $m=0.5$. }
\label{tab qnms2}
\begin{tabular}{p{0.24cm} p{0.24cm} p{0.2cm} p{3cm} p{2.8cm}}
\hline
 $\Lambda$& $l$	  & $n$  &  WKB & P\"{o}schl–Teller fitting\\ \hline 
 0.01& 1 & 0 &  		0.4128044-0.0371941$i$ & 0.412896 - 0.038664$i$ \\
 & 2 & 0 & 	0.5328532-0.0620077$i$	& 0.538972 - 0.063760$i$ \\
 &  & 1 &  0.4939282-0.1892938$i$ &	 0.538972 - 0.191281$i$	\\
 & 3 & 0 & 0.6926497-0.0722363$i$& 0.696234 - 0.072839$i$ \\
 &  &  1& 0.6725369-0.2198483$i$&  0.696234 - 0.218518$i$\\
 &  & 2 & 0.6356915-0.3758681$i$& 0.696234 - 0.364197$i$ \\
 \hline
 0.03 & 1 & 0 &   0.3451351-0.0516676$i$ & 0.345029 - 0.051558$i$\\
 & 2 & 0 &   	0.4611284-0.0607523$i$ &	 0.463409 - 0.061644$i$ \\
 &  & 1 & 	0.4452242-0.1808923$i$ & 0.463409 - 0.184934$i$ \\
 & 3 & 0 & 0.6009324-0.0658807$i$ & 0.602646 - 0.066297$i$\\
 &  &  1& 0.5907873-0.1980051$i$& 0.602646 - 0.198893$i$\\
 &  & 2 & 0.5706236-0.3312089$i$& 0.602646 - 0.331488$i$\\
 \hline
 0.05	&	1&	0&	0.2764212-0.0503552$i$&	 0.276227 - 0.050124$i$\\
 	&	2&	0& 	0.3772614-0.0541405$i$&	0.378049 - 0.054441$i$\\
 	&	&	1& 	0.3714364-0.1618169$i$& 0.378049 - 0.163326$i$\\
  & 3 & 0 & 0.4937238-0.0564043$i$&  0.49444 - 0.056596$i$\\
 &  &  1& 0.4893173-0.1691649$i$& 0.49444 - 0.169789$i$\\
 &  & 2 & 0.4803448-0.2818699$i$&  0.49444 - 0.282981$i$\\
 	\hline
0.07 & 1 & 0 & 	0.1962347-0.0403962$i$& 0.196117 - 0.040266$i$\\
 & 2 & 0 & 	0.2723391-0.041586$i$ & 0.272527 - 0.041641$i$\\
 &  & 1 &	0.2708491-0.1246458$i$& 0.272527 - 0.124925$i$ \\
& 3 & 0 & 0.3580321-0.0422989$i$&  0.358241 - 0.042351$i$\\
 &  &  1& 0.3567047-0.1268734$i$& 0.358241 - 0.127056$i$ \\
 &  & 2 & 0.3540054-0.2114033$i$&  0.358241 - 0.211759$i$\\
 \hline
 0.09	&	1&0	&   	 0.0649324-0.0146021$i$	& 0.064925 - 0.014592$i$\\
 	&	2&0	&	0.0914281-0.0146386$i$&	 0.091432 - 0.014638$i$\\
 	&	&1	&		0.0913917-0.0439152$i$& 0.091432 - 0.043914$i$\\
 & 3 & 0 & 0.1207429-0.0146611$i$& 0.120749 - 0.014662$i$ \\
 &  &  1& 0.1207028-0.0439836$i$ & 0.120749 - 0.043986$i$\\
 &  & 2 & 0.1206229-0.0733065$i$ & 0.120749 - 0.07331$i$\\
 	\hline
\end{tabular}
\end{table}
 In both the methods, for a fixed multipole and overtone number, the magnitudes of both real and the imaginary parts of the quasinormal frequencies decrease with increasing $L$. This shows that  the oscillation frequency and the damping rate decrease with increasing $L$. Further as the cosmological constant $\Lambda$ increases,  the oscillation frequency and the damping rate  decrease except for $l=1$. For $l=1$, the absolute value of the imaginary part increases at first and then decreases as $\Lambda$ increases. This implies that the damping rate is faster when $\Lambda$ is small and  is slower for increasing $\Lambda$.  For a fixed value of $l$ and increasing values of $n$,  the magnitude of the imaginary part increases indicating that higher values of $n$ lead to the higher damping rate.

\subsection{Dirac perturbation}

Using the effective potentials of  massless and massive Dirac field, the corresponding quasinormal frequencies are calculated by using the 3rd order WKB method and the P\"{o}schl–Teller fitting method. The quasinormal frequencies are computed for a range of values of $l$ and $n$, altering the Lorentz violation parameter $L$ and cosmological constant $\Lambda$. The results are presented in Tables \ref{tab qnmd1} and \ref{tab qnmd2}, respectively. The case of $L=0$ represents the quasinormal frequencies of the Schwarzschild-de Sitter black hole. We observe that the variations of real part of the massless and massive Dirac quasinormal frequencies show a similar behaviour to the case of scalar quasinormal frequencies. The oscillation frequency decreases with increasing $L$ and $\Lambda$. The magnitude of the imaginary parts of massless and massive Dirac quasinormal frequencies decrease with increasing $L$  implying  that the damping rate  decreases as $L$ increases. However, with varying $\Lambda$, we observe a difference in the damping rate of massless and massive Dirac field. For the massless Dirac field, the damping rate decreases with increasing $\Lambda$ but for the massive Dirac field,  we observed that at first the damping rate is increasing and later, the damping rate becomes slower  with increasing values of $\Lambda$. The results are summarised in Figs. \ref{qnmmassless}.

\begin{table*}[t]
\caption{Quasinormal frequencies for massless and massive Dirac perturbation  for different modes and for different values of the Lorentz violating parameter $L$ with fixed value of $\Lambda=0.05$,  $\omega=0.2$ and $\mu_*=1$. }
\label{tab qnmd1}
\begin{tabular}{p{0.3cm} p{0.3cm} p{0.3cm} p{3cm} p{3cm} p{3cm} p{3cm} }
\hline
&	&	& Massless Dirac	&	& Massive Dirac& 	 \\ \hline
 $L$ & $l$  & $n$ & WKB & P\"{o}schl–Teller fitting &	WKB		&P\"{o}schl–Teller fitting	  \\
 \hline
 0&1  &0  & 0.211104-0.0712303$i$  &  0.21139 - 0.071517$i$	&0.550085-0.0571941$i$&	0.549676 - 0.056798$i$	   \\
 & 2 &  0& 0.35511-0.0713101$i$ &  0.355831 - 0.071543$i$	&0.630096-0.0604895$i$&	0.630764 - 0.060686$i$		  \\
 &  &  1& 0.346379-0.215282$i$ & 0.355831 - 0.214629$i$	&0.625122-0.180314$i$&  0.630764 - 0.182058$i$			   \\
  & 3 &  0&  0.498348-0.0713335$i$ &  0.499045 - 0.071495$i$ 	&0.730631-0.0635102$i$&	 0.731594 - 0.063782$i$	  \\
 &  &  1& 0.492073-0.214684$i$ & 0.499045 - 0.214485$i$	&0.725174-0.190041$i$ &	0.731594 - 0.191348$i$	  \\
 &  &  2& 0.480457-0.359474$i$ & 0.499045 - 0.357475$i$	&0.713563-0.315373$i$&	0.731594 - 0.318913$i$	   \\
 \hline
 0.3& 1 &0  & 0.184487-0.0542644$i$ & 0.184525 - 0.054397$i$  &0.465727-0.0463371$i$&  0.465623 - 0.046203$i$		   \\
 &  2&  0& 0.309092-0.0543287$i$ &   0.309432 - 0.054441$i$	&0.534022-0.0479519$i$& 0.534271 - 0.047992$i$		   \\
 &    &  1& 0.304599-0.1634721$i$ &   0.309432 - 0.163324$i$ 	& 0.532266-0.143609$i$&	0.534271 - 0.143976$i$	  \\
  &   3 &  0& 0.433324-0.0543469$i$ &  0.433667 - 0.0544245$i$		&0.620533-0.0495298$i$&	0.620919 - 0.049613$i$	   \\
 &    &  1&  0.43013-0.163279$i$ & 	0.433667 - 0.163274$i$  & 0.618373-0.148444$i$&	0.620919 - 0.148842$i$		   \\
 &    &  2& 0.424043-0.272747$i$ & 0.433667 - 0.272123$i$	&0.613869-0.246988$i$&	0.620919 - 0.248069$i$		  \\
 \hline
 	0.5&	1&0	&	0.163694-0.0447171$i$&  0.163622 - 0.044778$i$ 	&0.407542-0.0394516$i$&		0.407541 - 0.039392$i$	  	\\
 	&	2&0	&	0.273776-0.0447631$i$&	0.273953 - 0.044825$i$	& 0.467719-0.0404312$i$&	0.467873 - 0.040446$i$		  \\
 	&	&	1&	0.271054-0.134514$i$&	0.273953 - 0.134476$i$	& 0.466752-0.121233$i$&  0.467873 - 0.121339$i$\\
 	 & 3 &  0&  		0.38364-0.0447759$i$ & 	0.383835 - 0.04482$i$	&0.544229-0.0414276$i$&	 0.544451 - 0.041465$i$	  \\
 &  &  1& 		0.381714-0.134436$i$ & 	0.383835 - 0.13446$i$		&0.543037-0.124239$i$ &	0.544451 - 0.124397$i$	  \\
 &  &  2& 		0.37793-0.224347$i$ &  	0.383835 - 0.2241$i$		&0.540597-0.20694$i$& 0.544451 - 0.207329$i$		   \\
 	\hline

0.8 & 1 &0  &  0.125515-0.0312317$i$ & 0.125366 - 0.031229$i$	&0.310096-0.0285932$i$&	0.310173 - 0.028585$i$		   \\
 &  2&  0&0.209531-0.0312509$i$  &	0.209562 - 0.031267$i$	&0.356134-0.0290143$i$&	0.356234 - 0.029021$i$		  \\
 &    &  1& 0.20851-0.093804$i$ &  0.209562 - 0.093803$i$	&0.355654-0.0870734$i$&	0.356234 - 0.087063$i$		  \\
  &   3 &  0& 0.293473-0.0312566$i$  & 0.293529 - 0.03127$i$  	& 0.41484-0.0294598$i$&	0.414949 - 0.029471$i$			  	 \\
 &    &  1& 0.292753-0.0937933$i$  &	 0.293529 - 0.093811$i$	&0.41433-0.0883972$i$&	0.414949 - 0.088415$i$			  	\\
 &    &  2&  0.29134-0.156387$i$& 	0.293529 - 0.156352$i$	&0.413324-0.147382$i$& 0.414949 - 0.088415$i$				  \\
 \hline
1 &	1&	0&	0.091184-0.0215068$i$&    0.091047 - 0.021489$i$  	& 0.228419-0.0199506$i$&	0.228512 - 0.019956$i$			  	\\
 &	2&0	&	0.152082-0.0215124$i$&  0.152068 - 0.021514$i$	& 0.261979-0.020165$i$	& 0.262063 - 0.020171$i$				  \\
 &	&	1&	0.151732-0.0645475$i$& 0.152068 - 0.064542$i$		& 0.261633-0.060536$i$& 0.262063 - 0.060514$i$			 	\\
 & 3 &  0&  		0.212958-0.0215143$i$ & 	0.212967 - 0.021517$i$	&0.30479-0.0203901$i$&				0.304864 - 0.020396$i$  \\
 &  &  1& 		0.212711-0.0645475$i$ & 	0.212967 - 0.064552$i$	&0.304481-0.0611974$i$ &				0.304864 - 0.061188$i$  \\
 &  &  2& 		0.212221-0.107592$i$ & 0.212967 - 0.107587$i$		&0.30389-0.102077$i$&				0.304864 - 0.101981$i$   \\
\end{tabular}
\end{table*}

\begin{table*}[t]
\caption{Quasinormal frequencies for massless and massive Dirac perturbationfor different modes and for different values of cosmological constant $\Lambda$ with fixed value of $L=0.2$, $\omega=0.2$ and $\mu_*=1$. }
\label{tab qnmd2}
\begin{tabular}{p{0.3cm} p{0.2cm} p{0.2cm} p{3cm} p{3cm} p{3cm} p{3cm} }
\hline
 &	&	& Massless Dirac	&	& Massive Dirac& 	 \\ \hline
 $\Lambda$ & $l$  & $n$ & WKB & P\"{o}schl–Teller fitting &	WKB		&P\"{o}schl–Teller fitting	  \\
 \hline
 0.01&1  &0  & 	0.2676723-0.0831291$i$	 &  			0.269457 - 0.083687$i$	 &	0.7917226-0.0373632$i$ &		0.790141 - 0.03669$i$\\
 & 2 &  0& 	0.4515244-0.0830115$i$	 &  0.453241 - 0.083386$i$	 &	0.8793448-0.0470078$i$&		0.8812 - 0.0476785$i$ 	\\
 &  &  1& 	0.4372108-0.2520782$i$ & 	0.453241 - 0.25016$i$	 &0.8605325-0.1324904$i$	&		0.8812 - 0.143036$i$ \\
  &   3 &  0& 		0.6341408-0.0829837$i$  & 	0.635578 - 0.083227$i$	 &	1.0011544-0.0580109$i$	&	1.00412 - 0.0586796$i$	 \\
 &    &  1& 		0.6236414-0.2505562$i$  &	 0.635578 - 0.249683$i$		 &
 0.9821202-0.172771$i$&		1.00412 - 0.176039$i$	\\
 &    &  2& 			 	0.6048075-0.4217087$i$& 	0.635578 - 0.416138$i$	&
 0.9418027-0.2841681$i$&		1.00412 - 0.293398$i$
 \\
 \hline
0.03 & 1 &0  & 		0.2340131-0.0721789$i$ & 	0.23487 - 0.072561$i$	 &0.6346851-0.050957$i$&		0.633988 - 0.050485$i$ \\
 &  2&  0& 	0.3936637-0.0721996$i$	 & 0.39468 - 0.072464$i$ 	 &0.7205787-0.0554771$i$&		0.721249 - 0.055645$i$	 \\
 &    &  1& 		0.3839023-0.2182295$i$ & 0.39468 - 0.217394$i$	 & 0.7143516-0.1646331$i$&	0.721249 - 0.166936$i$		\\
  &   3 &  0& 			 0.5524677-0.0722084$i$  & 	0.553358 - 0.072382$i$	&
  0.8299653-0.0598877$i$	 &	0.831136 - 0.060187$i$	\\
 &    &  1& 		0.5454138-0.2174616$i$  &	 0.553358 - 0.217147$i$ 	 &		
 0.8227519-0.1789212$i$	&	0.831136 - 0.180563$i$	\\
 &    &  2& 			0.5324119-0.3645758$i$& 		0.553358 - 0.361912$i$	 &
 0.8073638-0.2960143$i$&	0.831136 - 0.300938$i$	\\
 \hline
 0.05	&	1&0	&	0.1939083-0.0594601$i$	& 0.194018 - 0.059637$i$	 &0.4939637-0.049831$i$		&	0.493783 - 0.049638$i$ 		\\
 	&	2&0	&	0.3252364-0.0595324$i$	&	0.325682 - 0.059677$i$	 &0.5661596-0.0518845$i$	&	0.566493 - 0.051952$i$		\\
 	&	&	1&	0.319587-0.179288$i$	&	 0.325682 - 0.179033$i$	 &0.5637223-0.1552296$i$&		0.566493 - 0.155856$i$	\\
 	 &   3 &  0& 		0.4560861-0.0595529$i$  & 		0.456526 - 0.059652$i$	 &
 	 0.6573926-0.0538478$i$	&	0.65791 - 0.053972$i$	 \\
 &    &  1& 		0.4520578-0.1790004$i$  &	 0.456526 - 0.178959$i$			 &
 0.6544681-0.1613148$i$	&	0.65791 - 0.161917$i$	\\
 &    &  2& 			0.4444463-0.2992041$i$& 		0.456526 - 0.298265$i$	 &
 0.6483227-0.2682077$i$&		0.65791 - 0.269861$i$\\
 	\hline

0.07 & 1 &0  &  		0.1418895-0.0433121$i$ &  0.141541 - 0.043306$i$	 &0.3439504-0.0399012$i$		&	0.343997 - 0.03987$i$		 \\
 &  2&  0&		0.2372446-0.0433616$i$	 &	0.237294 - 0.0434$i$  &	0.3966156-0.0406048$i$		&	0.396776 - 0.040632$i$	\\
 &    &  1& 		0.2350082-0.1302559$i$ & 	0.237294 - 0.130201$i$ & 0.3957908-0.1217846$i$		&	0.396776 - 0.121898$i$	\\
  &   3 &  0& 		0.3324305-0.0433751$i$  & 	0.332544 - 0.043408$i$	 &	
  0.4634247-0.0412929$i$	 &	0.463624 - 0.041333$i$	\\
 &    &  1& 		0.3308513-0.1302059$i$  &		0.332544 - 0.130225$i$ & 		
 0.4624588-0.12386$i$	&	0.463624 - 0.124001$i$	\\
 &    &  2& 			0.3277977-0.2172132$i$& 			0.332544 - 0.217042$i$ &
 0.4604991-0.2063823$i$ &	0.463624 - 0.206668$i$	\\
 \hline
0.09 &	1&	0&		0.0482855-0.0147021$i$	&	0.0480726 - 0.014665$i$ &  0.1168829-0.0140893$i$	&		0.116956 - 0.0141002$i$			\\
 &	2&0	&	0.0804934-0.0146982$i$		& 	0.080439 - 0.0146909$i$  & 	0.1347123-0.0141721$i$		&			0.134772 - 0.0141801$i$\\
 &	&	1&	0.0804108-0.0440944$i$		 &  0.080439 - 0.044072$i$& 0.1344805-0.0425588$i$			&		0.134772 - 0.0425404$i$	\\
 &   3 &  0& 		0.1126999-0.0146981$i$  & 	0.11268 - 0.0146961$i$	 &	0.1574579-0.0142536$i$		 & 0.157506 - 0.014259$i$\\
 &    &  1& 		0.1126396-0.0440949$i$  &		0.11268 - 0.0440884$i$ & 		
 0.1572712-0.0427892$i$		 &  0.157506 - 0.042778$i$\\
 &    &  2& 			0.1125194-0.0734925$i$& 		0.11268 - 0.0734806$i$	 &
 0.1569317-0.0713949$i$	 & 0.157506 - 0.071297$i$\\
\end{tabular}
\end{table*}

\section{Spectrum and sparsity of Hawking radiation}\label{sparsity}
\begin{figure}[h!]
\centering
  \centerline{\includegraphics[height=160pt,width=230pt]{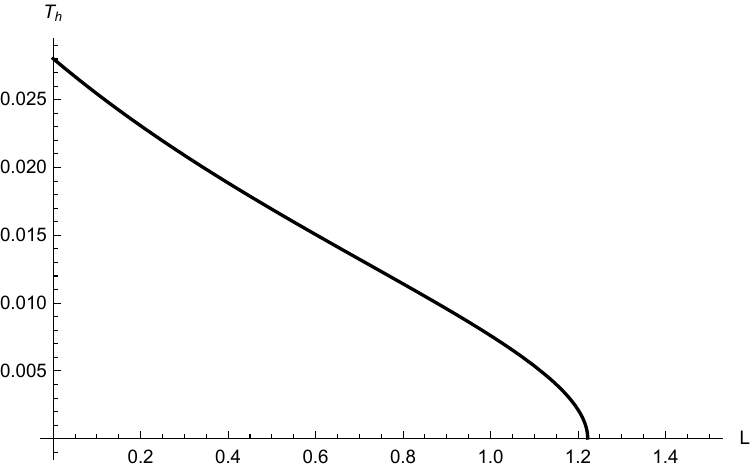}}
  \caption{Variation of Hawking temperature $T_h$ with respect to $L$ for fixed $M=1$ and $\Lambda=0.05$. } 
  \label{fig temperature}
\end{figure}

In this section, we investigate the effect of Lorentz invariance violation parameter $L$ on the spectrum and sparsity of Hawking radiation for the massless scalar and Dirac perturbation. First, we derive the Hawking temperature  as
\begin{align}\label{sparsity temp}
T_h=&\dfrac{1}{4\pi \sqrt{-g_{tt}g_{rr}}} \dfrac{d \, g}{dr}\vert_{r=r_h}\nonumber\\
=& \dfrac{1}{2\pi \sqrt{1+L}} \left(\dfrac{M}{r_{h}^2}-\dfrac{(1+L)r \Lambda}{3} \right).
\end{align}
When $L \rightarrow 0$, $T_{h}$ reduces to the original Hawking temperature of Schwarzschild-de Sitter black hole. To see the effect of $L$ on the Hawking temperature, we plot the Hawking temperature versus $L$ in Fig. \ref{fig temperature}. It is observed that the Hawking temperature decreases with increasing $L$.

\begin{figure}[h!]
\centering
  \centerline{\includegraphics[height=160pt,width=230pt]{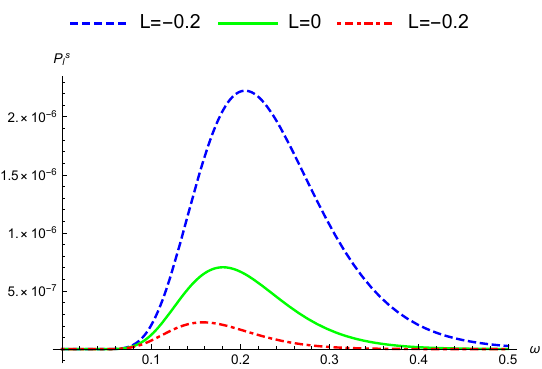}}
  \caption{Variation of Hawking temperature $T_h$ with respect to $L$ for fixed $M=1$ and $\Lambda=0.05$. } 
  \label{fig pls}
\end{figure}

\begin{figure}[h!]
\centering
  \centerline{\includegraphics[height=160pt,width=230pt]{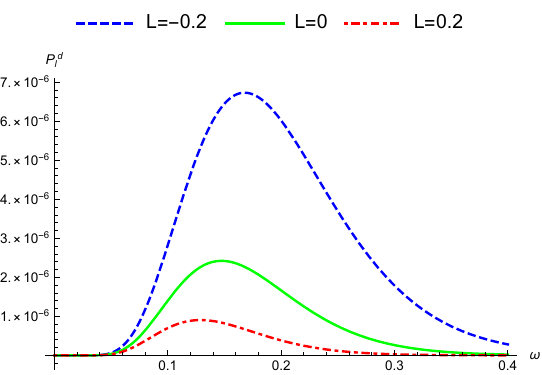}}
  \caption{Plot of the power spectrum of Dirac perturbation $P_{l}^d$ with $\omega$ for different values of $L$ for fixed $M=1$ and $\Lambda=0.05$. } 
  \label{fig pld}
\end{figure}

The total power emitted as Hawking radiation by a black hole at temperature $T_h$ with frequency $\omega$ in the momentum interval $d^3 \vec{k}$ is given by \cite{gray2016,miao2017}
\begin{align} \label{sparsity1}
\dfrac{dE(\omega)}{dt}\equiv P_{tot} =\sum\limits_l T(\omega) \dfrac{\omega}{e^{\omega/T_h}-1} \hat{k} \cdot \hat{n} \dfrac{d^3 k dA}{(2\pi)^3},
\end{align}
where  $\hat{n}$ is the unit vector normal to the surface element $dA$ and $T(\omega)$ is the greybody factor. For massless particles we have $\vert k \vert=\omega$, hence Eq. \eqref{sparsity1} reduces to
\begin{align}
P_{tot}=\sum\limits_l \int \limits_{0}^\infty P_l(\omega) d\omega,
\end{align}
where $P_l(\omega)$ is the power spectrum in the $l^{th}$ mode which is given by
\begin{align}
P_l(\omega)=\dfrac{A}{8 \pi^2} T(\omega) \dfrac{\omega^3}{e^{\omega/T_h}-1}.
\end{align}
Here, $A$ is a multiple of the horizon area. We consider  $A$ to be the horizon area as it will not affect the qualitative result \cite{miao2017}. From Figs. \ref{fig pls} and \ref{fig pld}, it is observed that the peak of the power spectrum for both the scalar and Dirac perturbation decrease  with increasing the value of $L$ and shift towards lower frequencies.

To discuss further the quantitative measure of the radiation emitted by black hole, we introduce a dimensionless parameter $\eta$ which defines the sparsity of Hawking radiation as \cite{gray2016,miao2017,hod2015,hod2016}
\begin{align}
\eta=\dfrac{\tau_{gap}}{\tau_{emission}}.
\end{align}
Here $\tau_{gap}$ is the average time gap between the emission of two successive Hawking radiation quanta, 
\begin{align}
\tau_{gap}=\dfrac{\omega_{max}}{P_{tot}},
\end{align}
where $\omega_{max}$ is the frequency at which the peak of the power spectrum occurs. $\tau_{emission}$ is the characteristic time for the emission of individual Hawking quantum,
\begin{align}
\tau_{emission} \geq \tau_{localisation}=\dfrac{2\pi}{\omega_{max}}.
\end{align}
where $\tau_{emission}$ is the characteristic time period of the emitted wave field with frequency $\omega_{max}$ to complete one cycle of oscillation. For $\eta\ll1$, the Hawking radiation flow is almost continuous and $\eta\gg 1$ implies a sparse Hawking radiation i.e. the time gap between the emission of successive Hawking quanta
emission is much larger than the time required for the emission of individual Hawking quantum. For different values of the Lorentz invariance violation parameter $L$, we present the numerical values of $\omega_{max}$, $P_{max}$, $P_{tot}$ and $\eta$ for massless scalar and Dirac perturbations  in Tables \ref{tab pls} and \ref{tab pld} respectively. For both the perturbations, it is found that increasing the parameter $L$ increases the sparsity of the black hole. The results show that the variation of sparsity is significant for both the perturbations. Thus, the Lorentz invariance violation has a significant impact  on the Hawking radiation.

\begin{table*}[]
\caption{Values of $\omega_{max}$, $P_{max}$, $P_{tot}$ and $\eta$ for massless scalar perturbation for various values of $L$ with fixed $l=1$ and $\Lambda=0.05$. }
\centering
\begin{tabular}{llllll}\label{tab pls}
 $L$				& 	0 	&	0.3	  &		0.5  &	0.8		&	1  \\ \hline
$ \omega^{s}_{max}$	&  	$0.180513$		&	0.147391  &  0.125887 & 	0.0915023	&   0.0640025	\\
$P^{s}_{max}	 $		&  $7.04852 \times 10^{-7}$ &	 $1.31583 \times 10^{-7} $ & 	$4.15876 \times 10^{-8}$	 &   $5.89315 \times 10^{-9}$	&	$1.01922 \times 10^{-9}$	\\
$P^{s}_{tot} $		&   $1.00885 \times 10^{-7}$	  &  $1.45918 \times 10^{-8}$ & 	$3.82098 \times 10^{-9}$	 & 	$3.77318\times 10^{-10}$ 	& 	$4.43875 \times 10^{-11}$	 \\
$\eta^s$		&	51405.8  &  236948  	& 	660099	 & 	$3531640 $	&	$14687700$	\\ \hline
\end{tabular}
\end{table*}




\begin{table*}[]
\caption{Values of $\omega_{max}$, $P_{max}$, $P_{tot}$ and $\eta$ for massless Dirac perturbation for various values of $L$ with fixed $l=1$, and $\Lambda=0.05$. }
\centering
\begin{tabular}{llllll}\label{tab pld}
 $L$				& 	0 	&	0.3	  &		0.5  &	0.8		&	1  \\ \hline
$ \omega^{d}_{max}$	&  	0.147388		&	$0.119928$ &   0.102153 &  0.073943	&	0.051571\\
$P^{d}_{max}	 $		&  $2.42139 \times 10^{-6}$ &	 	$5.53627 \times 10^{-7}$ & 	$2.01604 \times 10^{-7}$	 &  $3.58025 \times 10^{-8}$ 	&	$7.32984 \times 10^{-9}$	\\
$P^{d}_{tot} $		& $3.32113 \times 10^{-7}$	  & $5.83608 \times 10^{-8}$ 	 & 	$1.75155 \times 10^{-8}$	 & 	 $2.1515 \times 10^{-9}$	& 		$2.98164 \times 10^{-10}$ \\
$\eta^d$		&	10410.2  &  39223 	& 	94819.7	 & 	404457	&	1419630	\\ \hline
\end{tabular}
\end{table*}

\section{Null geodesic} \label{geodesic} 
\begin{figure}[h!]
\centering
  \centerline{\includegraphics[height=160pt,width=230pt]{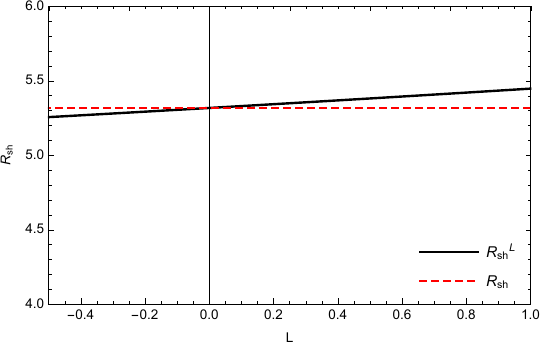}}
  \caption{Variation of $R_{shadow}$ with respect to $L$ for fixed $M=1$ and $\Lambda=0.05$. } 
  \label{fig shadow1}
\end{figure}

\begin{figure}[h!]
\centering
  \centerline{\includegraphics[height=200pt,width=200pt]{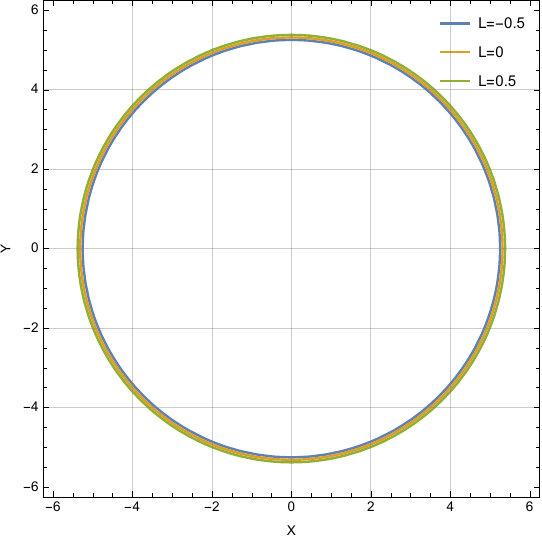}}
  \caption{Shadow in the celestial coordinate for different values of $L$ and fixed $M=1$ and $\Lambda=0.05$. } 
  \label{fig shadow2}
\end{figure}

In this section, we will study the null geodesics and the radius of the photon sphere for the  \\Schwarzschild-de Sitter-like black hole in bumblebee gravity model. Since the black hole is spherically symmetric, without loss of generality, one can assume the equatorial plane ($\theta=\pi/2$). The Lagrangian corresponding to the metric \eqref{metric} is defined as  
\begin{align}\label{lagrangian}
\La =\dfrac{1}{2} \left[f(r) \dot{t}^2	-\dfrac{1+L}{f(r)} \dot{r}^2 	-r^2 \dot{\phi}^2 \right].
\end{align}
The generalized momenta defined by $p_\mu=\dfrac{\partial \mathcal{L}}{\partial \dot{x}^\mu}=g_{\mu \nu}\dot{x}^\nu$ are derived from the Lagrangian as follows
\begin{align}
 \label{nullpt} p_{t}=& f(r) \dot{t}=E= \text{constant}, \\
 p_{r}=& \dfrac{1+L}{f(r)} \dot{r}, \\
\label{nullpphi} p_\phi=& -r^2 \dot{\phi}= -\Lb=\text{constant} .
\end{align}
Here $E$ and $\Lb$  are the constants of motion. Applying Eqs. \eqref{nullpt} and \eqref{nullpphi}, the Lagrangian in Eq. \eqref{lagrangian} is written as
\begin{align}\label{lagrangian2}
(1+L) \dot{r}^2+f(r) \left(\dfrac{L^2}{r^2}+\epsilon\right)-E^2=0.
\end{align}
Here, $\epsilon=2\La$. $\epsilon=0$ and 1 are associated with null and timelike  geodesics respectively. For photon, Eq. \eqref{lagrangian2} reduces to
\begin{align}\label{veff}
\dot{r}^2+ V_{eff}=0,
\end{align}
with $V_{eff}=\dfrac{f(r)}{(1+L)} \left(\dfrac{\Lb^2}{r^2}-\dfrac{E^2}{f(r)}\right)$.
For a circular null geodesic, the effective potential \eqref{veff} must satisfy the following three conditions,
\begin{align}\label{dveff}
V_{eff}=0, ~~~~~\dfrac{\partial V_{eff}}{\partial r_{ph}}=0,~~~~~
\dfrac{\partial^2 V_{eff}}{\partial r_{ph}^2}<0.
\end{align}
Using the above conditions \eqref{dveff}, we obtain
\begin{align}\label{radius}
2 f(r_{ph})-r_{ph} f'(r_{ph})=0.
\end{align}
On solving Eq. \eqref{radius}, one can obtain for our spherical the radius of the photon sphere as $r_{ph}=3M$. It is worth noticing that the Lorentz violation doesn't affect the radius of photon sphere. Further,  the shadow radius is calculated using the  celestial coordinate proposed as
\begin{align}
& X= \lim\limits_{r_0 \to \infty} \left(-r_{0}^2  \sin\theta_0 \frac{d\phi}{dr}\vert_{r_0,\theta_0} \right) \\
& Y= \lim\limits_{r_0 \to \infty} \left(r_{0}^2  \frac{d\phi}{dr}\vert_{r_0,\theta_0} \right),
\end{align}  where $r_0$ and $\theta_0$ are the positions of the observer at infinity. The shadow radius can be expressed as
\begin{align}\label{shadow}
R_{sh}^L=\sqrt{X^2+Y^2}=\dfrac{3\sqrt{3} M}{\sqrt{1-9(1+L)\Lambda M^2}}.
\end{align}
The shadow radius is modified due to the presence of $L$. In the limit $L\rightarrow 0$, we get shadow radius of Schwarzschild-de Sitter black hole ($R_{sh}$) \cite{kalita2023}. In Fig. \ref{fig shadow1}, we illustrate the variation of the shadow radius of the black hole with the Lorentz violation parameter $L$. It is observed that the shadow radius keeps on increasing with increasing $L$. However, the variation in shadow radius  is very small. A similar effect is also observed in the celestial coordinate in Fig. \ref{fig shadow2}.

\section{Eikonal quasinormal modes and photon geodesics connection}\label{eikonal}
 In this section, we study the relation between the null geodesics and quasinormal modes. At the eikonal limit, the real and imaginary parts of quasinormal frequencies of  spherically symmetric black holes are related to the frequency and instability time scale of unstable circular null geodesics  \cite{cardoso2009}. Thus, in the eikonal limit, the quasinormal frequencies can be obtained  by using the following expression \cite{devi2020,jusufi2020}
\begin{align}
\omega=\Omega l-i\left(n+\dfrac{1}{2}\right) \vert \lambda_L \vert,~~l\gg n,
\end{align}
where $\Omega$ is the angular velocity at the photon sphere,
\begin{align}\label{omega}
\Omega=\dfrac{\dot{\phi}}{\dot{t}}=\dfrac{\sqrt{f(r_{ph})}}{r_{ph}}.
\end{align}
The Lyapunov exponent $\lambda_L$ can be expressed as
\begin{align}\label{lyapunov}
\lambda_L=\sqrt{\dfrac{V_{eff}''}{2\dot{t}^2}}=\sqrt{\dfrac{f(r_{rp})}{\sqrt{2} (1+L) r_{rp}^2} \left[ r_{rp}^2 f''(r_{rp})-2 f(r_{rp})		\right]}.
\end{align}
Using Eqs. \eqref{omega} and \eqref{lyapunov}, we obtain the analytical form of the eikonal quasinormal frequencies of the Schwarzschild-de Sitter like black hole in Bumblebee gravity model
\begin{align}
\omega=&\dfrac{l~\sqrt{1-9(1+L)\Lambda M^2}}{3\sqrt{3}M}\nonumber\\
&-\dfrac{i}{3} \left(n+\dfrac{1}{2}\right) \sqrt{3\Lambda-\dfrac{1}{3 (1+L) M^2}}. 
\end{align}
The real part of the quasinormal mode and shadow radius are related by \cite{jusufi2020} 
\begin{align}
\omega_{Re}=\lim\limits_{l\gg1} \dfrac{l}{R_{sh}}.
\end{align}
which is accurate only in the eikonal limit having large values of $l$. However, this correspondence is not guaranteed for gravitational fields, as the link between the null geodesics and quasinormal modes is violated in the Einstein-Lovelock theory  in the eikonal limit \cite{konoplya2017}.

\section{Constraint on Lorentz violation parameter }\label{sec constraint}
\begin{figure}[h!]
\centering
  \centerline{\includegraphics[height=160pt,width=230pt]{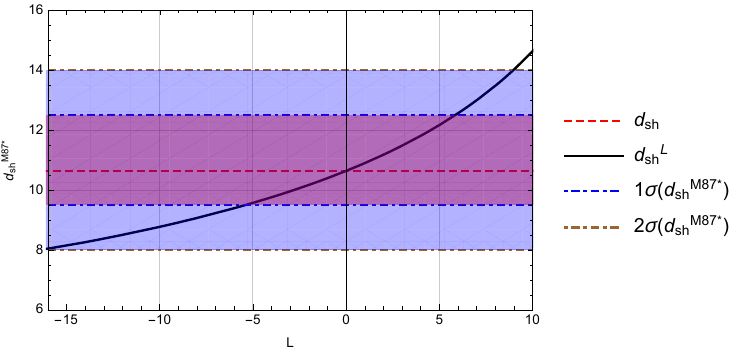}}
  \caption{Variation of $R_{shadow}$ with respect to $L$ for fixed $M=1$ and $\Lambda=0.05$. } 
  \label{fig m87}
\end{figure}

\begin{figure}[h!]
\centering
  \centerline{\includegraphics[height=160pt,width=230pt]{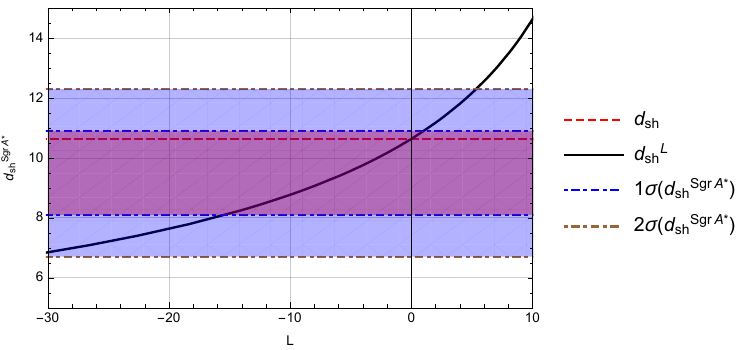}}
  \caption{Shadow in the celestial coordinate for different values of $L$ and fixed $M=1$ and $\Lambda=0.05$. } 
  \label{fig sgr}
\end{figure}

In this section, we shall constrain the Lorentz violation parameter $L$ using the EHT observation data. The shadow diameter of  M87* and Sgr A* have been realized through EHT observation \cite{akiyama2019a,akiyama2019b,akiyama2019c,akiyama2019d,akiyama2019e,akiyama2019f,akiyama2022}. The radial diameters of the shadow images for M87* and Sgr A* have been determined as $d_{sh}^{M87^*}= (11\pm 1.5)M$ and $d_{sh}^{Sgr{} A^*}= (9.5\pm 1.4)M$ respectively. The black hole shadow diameter of Schwarzschild-de Sitter-like black hole in Bumblebee gravity model ($d_{sh}$) can be obtained from Eq. \eqref{shadow}. It is important to notice from Eq. \eqref{shadow} that the constraint on the parameter $L$ depends on $M$ and $\Lambda$. The constraint on the parameter is obtained as
\begin{align}
L=\dfrac{1}{9M^2 \Lambda}-\dfrac{12}{D_{sh}^2 \Lambda}-1,
\end{align}
where $D_{sh}$ is the radial diameters of the shadow images obtained from the observational data. Figs. \ref{fig m87} and \ref{fig sgr}, we show the constraints on the Lorentz violation parameter $L$ from the observed shadow diameter of M87$^*$ and Sgr A$^*$ respectively. We consider $1\sigma$ and $2\sigma$ uncertainties to show the constraints. In Table \ref{constraint}, for the set of fixed parameter $M=1$ and $\Lambda=0.005$,   the corresponding boundary for the parameter $L$ is shown according to the experimental data.
\begin{table}[h!]
\caption{Variation in shadow with $1\sigma$ and $2\sigma$ ranges of $L$ based on the shadow diameter of M87$^*$ and Sgr A$^*$.}
\begin{tabular}{lcc}\label{constraint}
& \multicolumn{2}{l}{Constraint on $L$} \\
 & Lower bound  &  Upper bound \\ \hline
 $1\sigma$ from M87$^*$& -5.37058 &  5.86222 \\
$2\sigma$ from M87$^*$ &  -16.2778 &  8.97732 \\
$1\sigma$ from Sgr A$^*$ &  -15.3576 &  1.0219\\
$2\sigma$ from Sgr A$^*$ & -32.2418 &  5.35865 \\ \hline
\end{tabular}
\end{table}

\section{Conclusion}\label{conclusion}
In this paper, we investigate the scalar and Dirac perturbations in the background of Schwarzschild-de Sitter-like black hole in bumblebee gravity model. We examine the Klein-Gordon and Dirac equations, respectively, for scalar and
Dirac perturbations. The effective potentials of scalar and Dirac field are obtained by reducing the radial part of the solution into a  Schrödinger-like equation. To investigate the physical interpretations and the effect of parameters   $\Lambda$, $L$ and $l$ on the effective potentials,
we plot the effective potentials for different value of   parameters. The height of the effective potentials decrease monotonically  for increasing parameters $\Lambda$ and $L$ but the parameter $l$ causes the opposite effect. Further, the rigorous bounds on the greybody factors of scalar and Dirac field have also been calculated. It is found that the greybody factors  bound for both the scalar and Dirac field  increase with increasing the parameters $\Lambda$ and $L$. Conversely, when $l$ increases, the greybody factors bound decrease.  Thus, the higher  value of the effective potential, 
it is  more difficult for the waves to be transmitted and therefore, there exists a lower  bound of the greybody factor. Our findings are also consistent with the result obtained in quantum mechanics.
The quasinormal modes of scalar and Dirac perturbation are numerically computed using the WKB method of 3rd order and P\"{o}schl-Teller fitting method. For both the scalar and Dirac field, the oscillation frequency and the damping rate are found to be decreased with increasing values of $L$. Moreover, the analytical expression of quasinormal mode is also derived in eikonal limit.

  We study the Hawking spectrum and its sparsity. First, the
Hawking temperature is derived and the variation of the power spectrum with respect to $L$ is investigated using the greybody factors. For a fixed cosmological constant, the total Hawking radiation power emitted in each mode decreases with increasing $L$. The peak of the power spectrum also decreases and shifts towards lower frequencies. Studying the null geodesics and spherical photon orbits, the radius of the black hole shadow is also obtained. The shadow radius is found to be increased with increasing $L$. Moreover, We have used EHT observations data for $M87^*$ and $Sgr A^*$ to constrain parameters $L$ in Bumblebee gravity model.

Our findings show the significant influence of Lorentz violation parameter $L$ on several observables, such as effective potential, greybody factors, quasinormal modes, Hawking temperature, sparsity of Hawking radiation and black hole shadow. Our study will enhance the understanding of Bumblebee gravity model on the implications  on astrophysics significance. In near future, we will try to extend our work in studying the implications of Bumblebee gravity model in the optical observation such as Gravitational lensing.


\begin{thebibliography}{}
%
%

\bibitem{kostelecky1989a}V. A. Kostelecký, S. Samuel, Gravitational phenomenology in higher-dimensional theories and strings. Phys. Rev. D \textbf{40}, 1886
(1989)


\bibitem{casana2018} R. Casana, A. Cavalcante, F. P. Poulis, E. B. Santos, Exact Schwarzschild-like solution in a bumblebee gravity model. Phys. Rev. D \textbf{97},  104001 (2018)
\bibitem{mocioiu2000} I.
Mocioiu, M. Pospelov, R. Roiban, Low-energy limits on the antisymmetric tensor field background on the brane and on the non-commutative scale. Phys. Lett. B \textbf{489},
390 (2000)
\bibitem{carroll2001}S. M. Carroll, J. A. Harvey, V. A. Kostelecky, C. D. Lane,
 T. Okamoto, Noncommutative Field Theory and Lorentz Violation. Phys. Rev. Lett. \textbf{87}, 141601 (2001)

\bibitem{ferrari2007} A. F. Ferrari, M. Gomes, J. R. Nascimento, E.
Passos, A. Yu. Petrov,  A. J. da Silva, Lorentz violation in the linearized gravity. Phys. Lett. B \textbf{652},
174 (2007).

\bibitem{gambini1999}R. Gambini, J. Pullin, Nonstandard optics from quantum space-time. Phys. Rev. D \textbf{59}, 124021 (1999)

\bibitem{ellis2000}J. Ellis, N. E. Mavromatos,  D. V. Nanopoulos, Quantum-Gravitational Diffusion and Stochastic Fluctuations in the Velocity of Light.  Gen.
Relativ. Gravit.  \textbf{32}, 127 (2000)

\bibitem{kostelecky1989b}V. A. Kostelecký, S. Samuel, Spontaneous breaking of Lorentz symmetry in string theory. Phys. Rev. D \textbf{39}, 683
(1989)
\bibitem{kostelecky1989c}V. A. Kostelecký, S. Samuel, Phenomenological gravitational constraints on strings and higher-dimensional theories. Phys. Rev. Lett. \textbf{63}, 224 (1989)
\bibitem{kostelecky1991} V. A. Kostelecký, R. Potting, CPT and strings. Nucl. Phys. B \textbf{359}, 545
(1991)


\bibitem{jacobson2001}T. Jacobson, D. Mattingly, Gravity with a dynamical preferred frame. Phys. Rev. D \textbf{64}, 024028 (2001)
\bibitem{jacobson2004} T. Jacobson, D. Mattingly, Einstein-aether waves. Phys. Rev. D \textbf{70}, 024003 (2004)










\bibitem{colladay1997} D. Colladay, V. A. Kostelecký, CPT violation and the standard model. Phys. Rev. D \textbf{55}, 6760 (1997)



\bibitem{colladay1998} D. Colladay, V. A. Kostelecký, Lorentz-violating extension of the standard model. Phys. Rev. D \textbf{58}, 116002 (1998)

\bibitem{bluhm2005}R. Bluhm, V. A. Kostelecký, Spontaneous Lorentz violation, Nambu-Goldstone modes, and gravity. Phys. Rev. D \textbf{71}, 065008 (2005)

\bibitem{kostelecky2001} V. A. Kostelecký, R. Lehnert, Stability, causality, and Lorentz and CPT violation. Phys. Rev. D \textbf{63}, 065008 (2001)
\bibitem{kostelecky2004}V. A. Kostelecky, Gravity, Lorentz violation, and the standard model. Phys. Rev. D \textbf{69}, 105009 (2004)

\bibitem{bertolami2005} O. Bertolami, J. Paramos,  Vacuum solutions of a gravity model with vector-induced spontaneous Lorentz symmetry breaking. Phys. Rev. D \textbf{72} , 044001 (2005)
\bibitem{gullu2022}I. Gullu, A. Ovgun, Schwarzschild-like black hole with a topological defect in bumblebee gravity.  Ann. Phys. \textbf{436},  168721 (2022)

\bibitem{ding2022}C. Ding, X. Chen, X. Fu, Einstein–Gauss–Bonnet gravity coupled
to Bumblebee field in four dimensional spacetime. Nucl. Phys. B
\textbf{975}, 115688 (2022)

\bibitem{ovgun2019} A. Övgün, K. Jusufi, İ. Sakalli, Exact traversable wormhole solution in bumblebee gravity. Phys. Rev. D \textbf{99}, 024042 (2019)

\bibitem{ding2020}C. Ding, C. Liu, R. Casana, A. Cavalcate, Exact Kerr-like solution and its shadow in a gravity model with spontaneous Lorentz symmetry breaking. Eur. Phys. C \textbf{80}, 178
(2020)
\bibitem{ding2021}C. Ding, X. Chen, Slowly rotating Einstein-bumblebee black hole solution and its greybody factor in a Lorentz violation model*. Chin. Phys. C \textbf{45}, 025106 (2021)

\bibitem{maluf2022} R. V. Maluf, C. R. Muniz,  Comments on ``Greybody radiation and quasinormal modes of Kerr-like black hole in Bumblebee gravity model". Eur. Phys. J. C \textbf{82}, 94 (2022)

\bibitem{liu2023} W. Liu, X. Fang, J. Jing, J. Wang, QNMs of slowly rotating Einstein–Bumblebee black hole. Eur. Phys. J. C \textbf{83}, 83 (2023)

\bibitem{kanzi2022}S. Kanzi, I. Sakallı, Reply to “Comment on ‘Greybody radiation and quasinormal modes of Kerr-like black hole in Bumblebee gravity model’”. Eur. Phys. J. C \textbf{82}, 93 (2022)



\bibitem{ding2023} C. Ding, Y. Shi, J. Chen, Y. Zhou, C. Liu, Y. Xiao, Rotating BTZ-like black hole and central charges in Einstein-bumblebee gravity. Eur. Phys. J. C \textbf{83}, 573 (2023)

\bibitem{ding2023b}C. Ding, Y. Shi,     J. Chen,  Y. Zhou,      C. Liu, High dimensional AdS-like black hole and phase transition in Einstein-bumblebee gravity*. Chinese Phys. C \textbf{47}, 045102 (2023)



\bibitem{maluf2021} R. V. Maluf, J. C. S. Neves, Black holes with a cosmological constant in bumblebee gravity. Phys. Rev.  D \textbf{103}, 044002 (2021)
 
\bibitem{kanzi2019}S. Kanzi, I. Sakalli, GUP modified Hawking radiation in bumblebee gravity. Nucl. Phys. B \textbf{946}, 114703  (2019) 
 
 \bibitem{gomes2020}D. A. Gomes, R. V. Maluf, C. A. S. Almeida, Thermodynamics of Schwarzschild-like black holes in modified gravity models. Ann. Phys. \textbf{418},
168198 (2020).

\bibitem{sakalli2023} I. Sakalli, E. Yörük, Modified Hawking radiation of Schwarzschild-like black hole in
bumblebee gravity model. Phys. Scr. \textbf{98}, 125307 (2023)

\bibitem{karmakar2023} R. Karmakar, D. J. Gogoi, U. D. Goswami, Thermodynamics and shadows of GUP-corrected black holes with topological defects in Bumblebee gravity.  Phys. Dark Universe
\textbf{41}, 101249 (2023)
\bibitem{priyo2022} Y. P. Singh, T. I. Singh,
I. A. Meitei, A. K. Singh, Modified Hawking temperature of Kerr–Newman black
hole in Lorentz symmetry violation theory. Int. J. Mod. Phys. D \textbf{31}, 2250106 (2022)
\bibitem{onika2022} Y. O. Laxmi, T. I. Singh, I. A. Meitei, Modified entropy of Kerr-de Sitter black hole in Lorentz
symmetry violation theory. Gen. Relativ. Gravit. \textbf{54}, 77 (2022)

\bibitem{media2023} N. Media, Y. O. Laxmi, T. I. Singh, Fermions tunneling of Kerr–Newman–de Sitter
black hole in Lorentz violation theory. IJGMMP \textbf{20}, 2350217 (2023)

\bibitem{onika2023} Y. O. Laxmi, T. I. Singh, I. A. Meitei, Modified Hawking temperature and entropy of Kerr-de Sitter
black hole in Lorentz violation theory, Mod. Phys. Lett. A \textbf{38}, 2350089 (2023)
 \bibitem{onika2023b} Y. O. Laxmi, N. Media, T. I. Singh, P-v criticality of charged Reissner-Nordstrom-de Sitter black hole under the influence of Lorentz violation theory. Int. J. Mod. Phys. A \textbf{38}, 2350180 (2023)























\bibitem{ovgun2018}A. Ovgün, K. Jusufi, I. Sakalli, Gravitational lensing under the
effect of Weyl and bumblebee gravities: Applications of Gauss–
Bonnet theorem. Ann. Phys. \textbf{399}, 193–203 (2018).

\bibitem{li2020}Z. Li, G. Zhang, and A. Ovgün, Circular orbit of a particle and weak gravitational lensing. Phys. Rev. D \textbf{101}, 124058 (2020) 
 
\bibitem{carvalho2021} I.D.D. Carvalho, G. Alencar, W.M. Mendes, R.R. Landim, The
gravitational bending angle by static and spherically symmetric
black holes in bumblebee gravity. EPL \textbf{134}, 51001 (2021)

\bibitem{mangut2023}M. Mangut, H. Gürsel, S. Kanzi, I. Sakalli, Probing the Lorentz Invariance Violation via Gravitational Lensing and Analytical Eigenmodes of Perturbed Slowly Rotating Bumblebee Black Holes, Universe \textbf{2023}, 9(5), 225 (2023)
 
\bibitem{jha2021} S.K. Jha, A. Rahaman, Bumblebee gravity with a Kerr–Sen-like
solution and its Shadow. Eur. Phys. J. C \textbf{81}, 345 (2021).
 
 
\bibitem{wang2022} H.-M. Wang, S.-W. Wei, Shadow cast by Kerr-like black hole
in the presence of plasma in Einstein-bumblebee gravity. Eur.
Phys. J. Plus \textbf{137}, 571 (2022).
 
\bibitem{vagnozzi2023} S. Vagnozzi, et al., Horizon-scale tests of gravity theories and fundamental physics from the Event Horizon Telescope image of Sagittarius A*. Class. Quantum Grav. \textbf{40}, 165007 (2023) 
 
 
 
 \bibitem{yang2019} R.J. Yang, H. Gao, Y.G. Zheng, Q. Wu, Effects of Lorentz Breaking on the Accretion onto a Schwarzschild-like Black Hole. Commun. Theor. Phys. \textbf{71}, 568 (2019)
 \bibitem{kanzi2021}S. Kanzi, I. Sakalli, Greybody radiation and quasinormal
modes of Kerr-like black hole in Bumblebee gravity model.
Eur. Phys. J. C \textbf{81}, 501 (2021)
 \bibitem{uniyal2023}A. Uniyal, S. Kanzi, I. Sakalli, Some observable physical properties of the higher dimensional
dS/AdS black holes in Einstein-bumblebee gravity theory. Eur. Phys. J. C  \textbf{83}, 668 (2023)

\bibitem{oliveira2019}R. Oliveira, D. M. Dantas, V. Santos and C. A. S. Almeida, Quasinormal modes of bumblebee wormhole. Class. Quantum Grav. \textbf{36}, 105013 (2019)
 
 
\bibitem{oliveira2021} R. Oliveira, D.M. Dantas, C.A.S. Almeida, Quasinormal frequencies
for a black hole in a bumblebee gravity. EPL
\textbf{135}, 10003 (2021).

 
\bibitem{gogoi2022} D. J. Gogoi, U. D. Goswami, Quasinormalmodes and Hawking radiation sparsity of GUP corrected black holes in bumblebee gravity
with topological defects. JCAP \textbf{06}, 029 (2022).
 
\bibitem{chen2023} C. Chen, Q. Pan, J. Jing, Quasinormal modes of a scalar perturbation around a rotating BTZ-like black hole in Einstein-bumblebee gravity. Phys. Lett. B \textbf{846}, 138186 (2023)
 
 \bibitem{lin2023}R. H. Lin, R. Jiang, X. H. Zhai, Quasinormal modes of the spherical bumblebee black holes with a
global monopole. Eur. Phys. J. C  \textbf{83}, 720 (2023)
 




 \bibitem{hawking1975}S.W. Hawking, Particle creation by black holes. Commun. Math. Phys. \textbf{43}, 199 (1975)
 
 \bibitem{fernando2005}S. Fernando, Greybody factors of charged dilaton black holes in 2 + 1 dimensions. Gen. Relativ. Gravit. \textbf{37}, 461 (2005)
\bibitem{kim2008} W. Kim,  J.J. Oh, Greybody Factor and Hawking Radiation of Charged Dilatonic Black Holes. J. Korean Phys. Soc. \textbf{52}, 986 (2008) 
 
 
 
 \bibitem{parikh2000}M. K. Parikh, F. Wilczek, Hawking Radiation As Tunneling. Phys. Rev. Lett. \textbf{85}, 5042 (2000)
 
 \bibitem{konoplyaa2020}R. A. Konoplyaa, A. F. Zinhailoa, Grey-body factors and Hawking radiation of black holes in 4D Einstein-Gauss-Bonnet gravity. Phy. Lett. B \textbf{810}, 135793 (2020) 
 
 
 \bibitem{visser1999} M. Visser,  Some general bounds for one-dimensional scattering. Phys. Rev. A \textbf{59}, 427 (1999)

\bibitem{boonserm2008} P. Boonserm, M. Visser, Bounding the Bogoliubov coefficients.  Ann. Phys. \textbf{323}, 2779 (2008)


\bibitem{sakalli2022} I. Sakalli, S. Kanzi, Topical Review: greybody factors and quasinormal modes for black holes invarious theories - fingerprints of invisibles, Turk. J. Phys., \textbf{46}, 2 (2022)
\bibitem{boonserm2008b}P. Boonserm, M. Visser, Bounding the greybody factors for Schwarzschild black holes. Phys. Rev. D 78, 101502 (2008)

\bibitem{ngampitipan2013}T. Ngampitipan, P. Boonserm, 
Bounding the Greybody Factors for Non-rotating Black Holes. Int. J. Mod. Phys. D 22, 1350058
(2013)
\bibitem{boonserm2013} T. Ngampitipan, P. Boonserm, Bounding the greybody factors for the Reissner-Nordström black holes. J. Phys.: Conf. Ser. \textbf{435}, 012027 (2013)
\bibitem{boonserm2014a}P. Boonserm, T. Ngampitipan, M. Visser, Bounding the greybody factors for scalar field excitations on the Kerr-Newman spacetime. J. High Energ. Phys. \textbf{2014}, 113 (2014)
\bibitem{boonserm2014b}P. Boonserm, A. Chatrabhuti, T. Ngampitipan, M. Visser, Greybody factors for Myers–Perry black holes. J. Math.
Phys. 55, 112502 (2014)
\bibitem{boonserm2018}P. Boonserm, T. Ngampitipan, P. Wongjun, Greybody factor for black holes in dRGT massive gravity, Eur. Phys. J. C \textbf{78} 492 (2018) 

\bibitem{badawi2024} Al-Badawi, S. K. Jha, A. Rahaman, The fermionic greybody factor and quasinormal modes of hairy
black holes, as well as Hawking radiation’s power spectrum and
sparsity, Eur. Phys. J. C \textbf{84}, 145 (2024)

 
 
\bibitem{lenzi2023}M. Lenzi, C. F. Sopuerta, Black hole greybody factors from Korteweg–de Vries integrals: Computation, Phys. Rev. D \textbf{107}, 084039 (2023)


\bibitem{gray2016}F. Gray, S. Schuster, A. Van-Brunt, M. Visser, The Hawking cascade
from a black hole is extremely sparse. Class. Quantum Gravity
\textbf{33}, 115003 (2016)
\bibitem{miao2017}Y.-G. Miao, Z.-M. Xu, Hawking radiation of five-dimensional
charged black holes with scalar fields. Phys. Lett. B \textbf{772}, 542
(2017)
\bibitem{hod2015}S. Hod, The Hawking evaporation process of rapidly-rotating
black holes: an almost continuous cascade of gravitons. Eur. Phys.
J. C \textbf{75}, 329 (2015)
\bibitem{hod2016}S. Hod, The Hawking cascades of gravitons from higher dimensional
Schwarzschild black holes. Phys. Lett. B \textbf{756}, 133
(2016)
 
 
 
 \bibitem{konoplya2011}R. A. Konoplya, A. Zhidenko, Quasinormal modes of black holes: From astrophysics to string theory. Rev. Modern Phys. \textbf{83}, 793
(2011)
 
\bibitem{detweiler1980}S. L. Detweiler, Black holes and gravitational waves. III-The resonant frequencies of rotating holes. Astrophys. J. \textbf{239}, 292 (1980) 
 
 
 \bibitem{abbott2016}B. P. Abbott et al., Observation of Gravitational Waves from a Binary Black Hole Merger. Phys. Rev. Lett. \textbf{116}, 061102 (2016)

\bibitem{schutz1985}B. F. Schutz, C. M. Will, Black hole normal modes-A semianalytic approach. Astrophys. J. \textbf{291}, L33 (1985)

 \bibitem{iyer1987b} S. Iyer, Black-hole normal modes: A WKB approach. II. Schwarzschild black holes. Phys. Rev. D \textbf{35}, 3632 (1987)
 
\bibitem{iyer1987a} S. Iyer, C. M. Will, Black-hole normal modes: A WKB approach. I. Foundations and application of a higher-order WKB analysis of potential-barrier scattering. Phys. Rev. D \textbf{35}, 3621 (1987) 
 
 
 
\bibitem{wahlang2017}W. Wahlang, P. A. Jeena, S. Chakrabarti, Quasinormal modes of scalar and Dirac perturbations of Bardeen de Sitter black holes,  Int. J. Mod. Phys. D \textbf{26}, 1750160 (2017)
 
 
 
 \bibitem{leaver1985}E. W. Leaver, An analytic representation for the quasi-normal modes of Kerr black holes. Proc. Roy. Soc. Lond. A \textbf{402}, 285 (1985).
 
 \bibitem{ferrari1984}V. Ferrari, B. Mashhoon, New approach to the quasinormal modes of a black hole.
Phys. Rev. D \textbf{30}, 295 (1984)
 
 
 
 
 
 
\bibitem{cardoso2003} V. Cardoso, J. P. S. Lemos, Quasinormal modes of the near extremal Schwarzschild-de Sitter black hole. Phys. Rev. D \textbf{67}, 084020 (2003).
 
 
 
\bibitem{panotopoulos2018} G. Panotopoulos, Electromagnetic quasinormal modes of the nearly-extremal higher-dimensional Schwarzschild–de Sitter black hole. Mod. Phys. Lett. A \textbf{33}, 1850130 (2018)
 
\bibitem{rincon2018}A. Rincon, G. Panotopoulos, Greybody factors and quasinormal modes for a nonminimally coupled scalar field in a cloud of strings in (2+1)dimensional background. Eur. Phys. J. C \textbf{78}, 858 (2018). 
 
 
 
 
\bibitem{cardoso2009}V. Cardoso, A.S. Miranda, E. Berti, et. al, Geodesic stability, Lyapunov exponents, and quasinormal modes, Phys. Rev. D \textbf{79}, 064016 (2009). 
 
 
 
 
 
 
 \bibitem{akiyama2019a} K. Akiyama et al. (Event Horizon Telescope), First M87 event
horizon telescope results. I. The shadowof the supermassive black
hole. Astrophys. J. Lett. \textbf{875}, L1 (2019)
\bibitem{akiyama2019b} K. Akiyama et al. (Event Horizon Telescope), First M87 event
horizon telescope results. II. Array and instrumentation. Astrophys.
J. Lett. \textbf{875}, L2 (2019)
\bibitem{akiyama2019c}K. Akiyama et al. (Event Horizon Telescope), First M87 event
horizon telescope results. III. Data processing and calibration.
Astrophys. J. Lett. \textbf{875}, L3 (2019)
\bibitem{akiyama2019d} K. Akiyama et al. (Event Horizon Telescope), First M87 event
horizon telescope results. IV. Imaging the central supermassive
black hole. Astrophys. J. Lett. \textbf{875}, L4 (2019)
\bibitem{akiyama2019e} K. Akiyama et al. (Event Horizon Telescope), First M87 event
horizon telescope results. V. Physical origin of the asymmetric
ring. Astrophys. J. Lett. \textbf{875}, L5 (2019)
\bibitem{akiyama2019f} K. Akiyama et al. (Event Horizon Telescope), First M87 event
horizon telescope results. VI. The shadow and mass of the central
black hole. Astrophys. J. Lett. \textbf{875}, L6 (2019)
\bibitem{akiyama2022} K. Akiyama et al. (Event Horizon Telescope), First Sagittarius A*
Event Horizon Telescope Results. I. The shadow of the supermassive
black hole in the Center of theMilkyWay. Astrophys. J. Lett.
\textbf{930}, L12 (2022)
 
 
 
 
 
 
 
 
 
 
 
 \bibitem{rahman2012} M. A. Rahman, M. I. Hossain, Hawking radiation of Schwarzschild–de Sitter black hole by Hamilton–Jacobi method. Phys. Lett. B \textbf{712}, 1 (2012)
 
 
\bibitem{fulling1989} S. A. Fulling, Aspects of Quantum Field Theory in Curved Spacetime (Cambridge University Press, 1989)







\bibitem{toshmatov2017}B. Toshmatov, Z. Stuchlık, Slowly decaying resonances of massive scalar fields around
Schwarzschild-de Sitter black holes. Eur. Phys. J. Plus \textbf{132} 324 (2017)
\bibitem{zinhailo2024}A. F. Zinhailo, Exploring unique quasinormal modes of a massive scalar field in brane-world scenarios. Phys. Lett. B \textbf{853}  138682 (2024)


\bibitem{zhidenko2004} A. Zhidenko, Quasi-normal modes of Schwarzschild–de Sitter black holes. Class. Quantum Grav. \textbf{21}, 273 (2004)

\bibitem{newman1962}E. Newman, R. Penrose, An Approach to Gravitational Radiation by a Method of Spin Coefficients. J. Math. Phys. \textbf{3}, 566 (1962)
\bibitem{chandrasekhar} S. Chandrasekhar, The Mathematical theory of Black holes (Oxford University Press, 1983)

\bibitem{goldberg1967}J. N. Goldberg, A. J. Macfarlane, E. T. Newman, F. Rohrlich, E.C.G. Sudarshan Crossmark, Spin-s Spherical Harmonics and $\eth$. J. Math. Phys. \textbf{8}, 2155 (1967)


%



\bibitem{lambiase2023}G. Lambiase, L. Mastrototaro, R. C. Pantig,
 A. Övgün, Probing Schwarzschild-like black holes
in metric-affine bumblebee gravity with
accretion disk, deflection angle,
greybody bounds, and neutrino
propagation. J. Cosmol. Astropart. Phys. \textbf{12}, 026 (2023)

\bibitem{jha2024}S. K. Jha, A. Rahaman, Quasinormal modes, and different aspects of Hawking radiation
within the metric-affine bumblebee gravity framework. Nucl. Phys. B \textbf{1002}, 116536 (2024)


\bibitem{jha2023b}S. K. jha, Shadow quasinormal modes, greybody bounds, and Hawking sparsity of loop quantum gravity motivated non-rotating black hole. Eur. Phys. J. C \textbf{83}, 952 (2023)
\bibitem{boonserm2023x} P. Boonserm, S. Phalungsongsathit, K. Sansuk, P. Wongjun. Greybody factors for massive scalar field emitted from black holes in dRGT massive gravity. Eur. Phys. J. C \textbf{83}, 657 (2023)

\bibitem{kalita2023}S. Kalita, P. Bhattacharjee, Constraining spacetime metrics within and outside general relativity through the Galactic Center black hole (SgrA*) shadow, Eur. Phys. J. C  \textbf{83}, 120 (2023)
\bibitem{jusufi2020} K. Jusufi, Quasinormal modes of black holes surrounded by dark matter and their connection with the shadow radius, Phys. Rev. D \textbf{ 101}, 084055 (2020)
\bibitem{devi2020}S. Devi, R. Roy, S. Chakrabarti, Quasinormal modes and greybody factors of the novel four dimensional Gauss–Bonnet black holes in asymptotically de Sitter space time: scalar, electromagnetic and Dirac perturbations. Eur. Phys. J. C \textbf{80}, 760 (2020)
\bibitem{konoplya2017}R. A. Konoplya and Z. Stuchlk, Are eikonal quasinormal modes linked to the unstable circular null geodesics?, Phys. Lett. B \textbf{771},  597 (2017)

\end{thebibliography}
\end{document}